\documentclass[aps,prb,twocolumn,groupedaddress,nofootinbib]{revtex4-1}
\pdfoutput=1
\usepackage{amsmath}
\usepackage{amssymb}
\usepackage{natbib}
\usepackage{graphicx,xcolor}
\usepackage{bm,hyperref}

\usepackage{amsmath,bm}% this includes align environment
\begin{document}

\title{Superfluid-insulator transition and the BEC-BCS crossover in the Rashba moat band}

\author{Hassan Allami}
\affiliation{Department of Physics and Astronomy, The University of Utah, Salt Lake City, UT 84112, USA}
\author{O. A. Starykh}
\affiliation{Department of Physics and Astronomy, The University of Utah, Salt Lake City, UT 84112, USA}
\author{D. A. Pesin}
\affiliation{Department of Physics and Astronomy, The University of Utah, Salt Lake City, UT 84112, USA}
\affiliation{Department of Physics, University of Virginia, Charlottesville, VA 22904, USA}
\date{\today}

\begin{abstract}
We study the superconducting transition in a two-dimensional electron gas with strong Rashba spin-orbit coupling. We assume low electron density, such that only the majority spin band participates in the transition. We show that the superconducting transition follows either the Bose-Einstein condensation (BEC), or the Bardeen-Cooper-Schrieffer (BCS) scenarios, depending on the position of the chemical potential with respect to the bottom of the majority band, and the strength of the Coulomb repulsion between electrons. Hence, the BEC-BCS crossover in this system can be driven either by the change in the chemical potential, or the distance to a gate.
\end{abstract}

%\pacs{74.20.Fg}
%\keywords{Moat band, superconductivity, BCS model}

\maketitle

\section{Introduction \label{sec:intro}}

BEC-BCS crossover between the limits of Bose-Einstein condensation of tightly bound two-electron molecules and superconducting transition of widely overlapping pairs
lies at the heart of the current understanding of fermionic superfluidity. The subject has received an extensive theoretical and experimental attention in variety of systems including 2D systems with spin-orbit-coupling \cite{leggett1980diatomic,bec-bcs_so_1,bec-bcs_2D,fermion_sf_bec-bcs, bcs-bec-soc}, culminating in the observation of the crossover phenomenon in dilute alkali gases\cite{bec_bcs_1, bec_bcs_2}.
In this work, we study the BEC-BCS in a two-dimensional (2D) electron gas with low carrier density, realized in a 2D heterostructure with structural inversion asymmetry, and the ensuing Rashba spin-orbit coupling. Systems with similar conditions have been previously reported \cite{soc_HgTe, soc_Bi-surf, soc_exp_1, soc_exp_2, soc_exp_3, annular_FS}.

There have been a number of theoretical studies of fermionic superfluidity in systems with a single particle spectrum having a minimum on a line in 2D, or a surface in 3D  -- the ``moat'' \cite{sc_soc2d_rashba_gorkov,sc_soc_cappelluti,Chamon, bound-state_soc_1, bound-state_soc_2, bound-state_soc_3}.
The former option is realized in the lower spin subband of a Rashba 2DEG, and is considered in this work. The main attractive feature of such systems is their effectively one-dimensional (1D) thermodynamics,
brought about by the diverging 1D density of states (DoS) near the bottom of the moat band. The 1D DoS of moat bands makes it easier to form two-particle bound states, and to form Bose-Einstein condensates.
For instance, in 2D systems, which are the primary interest in the present work, the critical temperature shows a power-law dependence on the BCS coupling constant, as opposed to the exponentially small result for the usual parabolic dispersion in 2DEGs with constant DoS \cite{sc_soc_cappelluti}. This gives the proposal a higher chance of experimental observation.

In this paper, we study the competition between BCS attraction and Coulomb repulsion in the phase diagram of a 2DEG with a moat band. The repulsion is necessarily present in an electronic system, and the lack of screening in a 2D system with low carrier density makes Coulomb effects strong. We show that electrostatic repulsion between electrons, even when partially screened by an external gate, is of qualitative importance at the superconducting transition. Strong electrostatic repulsion induces an interesting BCS-BEC crossover in the presence of a shallow Fermi sea in the moat band.

The rest of the paper is organized as follows. In section~\ref{sec:model} we construct the model of a 2DEG with Rashba SOC. Assuming a low density limit next we introduce an attractive and a repulsive interaction term projected on the lower moat-like band. In section~\ref{sec:pairs} we treat the problem by considering a single pair wavefunction and showing that it can describe many features of superfluid-insulator transition including the effects of Coulomb repulsion. Section~\ref{sec:BCS} extends the problem to the case of a shallow ring-like Fermi sea. There we study the BEC-BCS crossover and construct the phase diagram of the system in the absence and presence of Coulomb repulsion.
Finally we provide a brief summary of the results in Section~\ref{sec:conclusion}.
Our main findings are summarized in phase diagrams in Figures \ref{fig:PDneutral} and \ref{fig:PDcoulomb}.

\section{Model of a 2DEG with a moat band \label{sec:model}}
\begin{figure}[h]
\includegraphics[width=0.8\columnwidth]{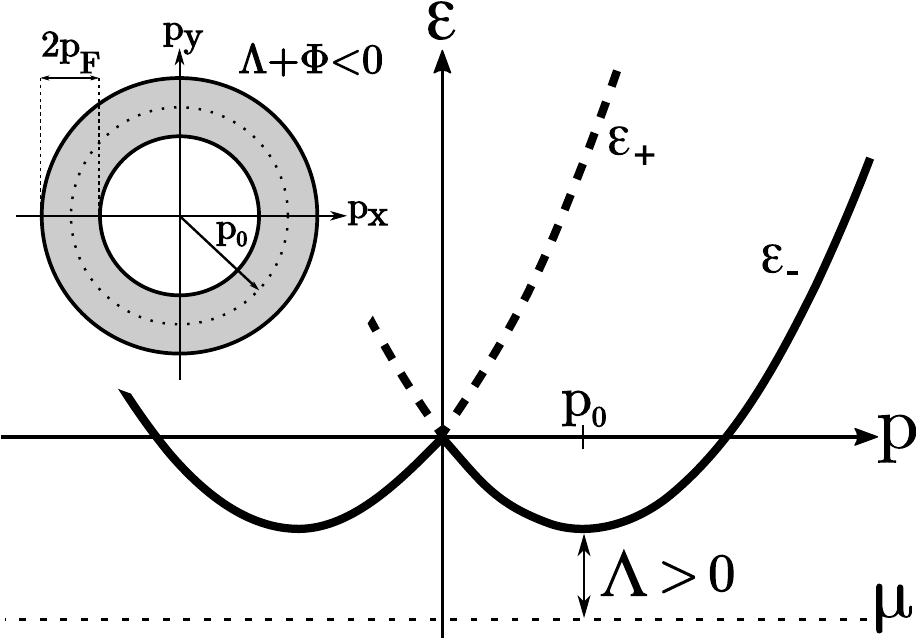}
\caption{Main: Side-view of the two 2D Rashba spin-orbit split bands for $\Lambda>0$ situation.
The insert shows top view of the ring-like Fermi surface for $\Lambda + \Phi < 0$ case, see Eq.\eqref{eq:pf} and related to it discussion.}
\label{fig:bands}
\end{figure}

For a system with the Rashba spin-orbit coupling that stems from structural inversion asymmetry, the single particle Hamiltonian in the momentum space has the following form:
\begin{equation}
H_0 = \kappa p^2 + \alpha \hat{\bm z}\cdot (\bm {p}\times\bm {\sigma}),
\label{eq:H0}
\end{equation}
 where $p$ is the particle's momentum, and $\alpha$ is the characteristic velocity that determines the strength of the spin-orbit coupling. The two eigen-bands of $H_0$ are given by
\begin{equation}
\varepsilon_\pm = \kappa p^2 \pm \alpha p = \kappa (p \pm p_0)^2 - E_R,
\label{eq:bands}
\end{equation}
where we defined the Rashba energy, $E_R = \alpha^2/(4\kappa)$.
The corresponding dispersions are plotted in Fig~\ref{fig:bands}.

The lower (``moat'') band, $\varepsilon_-$, has a minimum at a circle in momentum space, which is centered at the origin, and has the radius of $p_0 = \alpha / 2\kappa$. The energy minimum is given by $\varepsilon_-(p_0) = - E_R$. Therefore, the single-particle energy of the moat-shaped band,
measured from the chemical potential, can be written as
\begin{equation}
\xi_{\bm p} = \kappa (p-p_0)^2 +\Lambda,
\label{eq:xi}
\end{equation}
where $\Lambda$ is the minimum energy of the moat band measured from the chemical potential (see Fig.~\ref{fig:bands}).

The single-particle wave functions that correspond to the bands of Hamiltonian~\eqref{eq:H0} are represented by spinors, in which the spin orientation is determined by a particle's momentum due to the presence of the spin-orbit coupling. The spinors for the $``\pm"$ bands have the following form in the basis of $\sigma_z$ eigenstates:
\begin{align}
& \chi_+ = \frac{1}{\sqrt{2}}[1 , ie^{i\theta_{\bm p}}]
&& \chi_- = \frac{1}{\sqrt{2}}[i  e^{-i\theta_{\bm p}},1]
\label{eq:spins}
\end{align}
where $\theta_{\bm p}$ is the angle momentum ${\bm p}$ makes with the $\hat{x}$-axis, hence $e^{i\theta_{\bm p}} = \hat{p}_x + i \hat{p}_y$.

Now we consider the general form of density-density interaction
\begin{equation}
H'_{int} = \frac{1}{2V}\sum_{\sigma\sigma'}\sum_{{\bm {pp'q}}}U_{\bm q}
c_{{\bm {p+q}},\sigma}^\dagger c_{{\bm {p'-q}},\sigma'}^\dagger c_{{\bm {p'}},\sigma'}^{}
c_{{\bm p},\sigma}^{},
\label{eq:general int}
\end{equation}
where $U_{\bm q}$ is the Fourier transform of the interaction potential, and $c_{{\bm p},\sigma}$ is the annihilation operator of a particle with momentum ${\bm p}$ and spin $\sigma$. We are interested in the low-density case, in which the chemical potential lies far below $\varepsilon_+$ band bottom, such that this band's influence on the particle dynamics is insignificant. This allows to project the interaction Hamiltonian on the lower ($``-"$) band with the help of Eq.~\eqref{eq:spins}:
\begin{equation}
H_{int} = \frac{1}{8V}\sum_{{\bm {pp'q}}} U_{\rm eff}({\bm {p, p', q}})
c_{{\bm {p+q}}}^\dagger c_{{\bm {p'-q}}}^\dagger c_{{\bm {p'}}}^{} c_{{\bm p}}^{},
\label{eq:general int proj}
\end{equation}
with
\begin{equation}
U_{\rm eff}({\bm {p, p', q}}) = U_{\bm q}
(1+e^{-i\theta_{{\bm {p'}}}+i\theta_{{\bm {p'-q}}}})(1+e^{-i\theta_{{\bm p}}+i\theta_{{\bm {p+q}}}}).
\label{eq:U_eff}
\end{equation}

In what follows we are interested in the BEC-BCS crossover in the system of Fermions described by Hamiltonians~\eqref{eq:H0} and \eqref{eq:general int proj}. Specifically, we restrict ourselves to the case where the interaction Hamiltonian, that is the amplitudes $U_{\bm q}$, contain attraction in the Cooper channel, and repulsion in the Coulomb channel.

The Cooper channel of the problem is described by a contact attractive interaction of $U_{\bm q}\to-2g$ only between electrons with opposite momenta $\bm{p'} = -\bm{p}$.
We assume that this attractive interaction $g$ stems from the usual electron-phonon interaction, its specific value being determined by the details of the material hosting two-dimensional electron gas.
Note that spins of these electrons are also opposite to each other. In the Cooper channel, the phase factor in \eqref{eq:U_eff} reduces to $(1+e^{-i\theta_{{\bm p}}+i\theta_{\bm {p+q}}})^2$;
however since $c_{\bm p} c_{-\bm{p}} = - c_{-\bm{p}} c_{\bm{p}}$ even in momentum terms vanish and the pairing interaction can be written as
\begin{equation}
H_p = -\frac{g}{2V}\sum_{{\bm pp'}}
e^{-i\theta_{{\bm p}} +i\theta_{{\bm {p'}}}}
c_{{\bm {p'}}}^\dagger c_{{\bm {-p'}}}^\dagger c_{{\bm {-p}}}^{} c_{{\bm p}}^{}.
\label{eq:attraction}
\end{equation}

For the Coulomb channel however, we work with full interaction form presented in \eqref{eq:U_eff}. The significance of the phase factor, as we show in the rest of the paper, shall depend on the external screening mechanism that we introduce below.
Therefore, the repulsive part of the Hamiltonian maintains the general form in \eqref{eq:general int proj} and \eqref{eq:U_eff}.
\begin{equation}
H_c = \frac{1}{8V}\sum_{{\bm {pp'q}}} U_{\rm eff}({\bm {p, p', q}})
c_{{\bm {p+q}}}^\dagger c_{{\bm {p'-q}}}^\dagger c_{{\bm {p'}}}^{} c_{{\bm p}}^{}.
\label{eq:Coulomb_Ham}
\end{equation}

\begin{figure}[h]
\includegraphics[width=0.8\columnwidth]{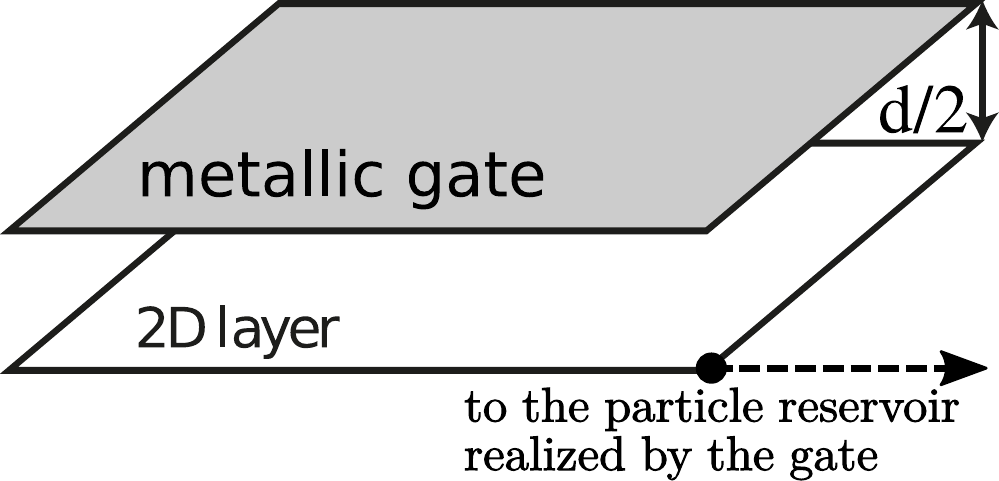}
\caption{Schematic for the gated 2D layer system.}
\label{fig:setup}
\end{figure}

The long-range nature of the Coulomb interaction brings up the issue of screening in a system of charged Fermions. In the two-dimensional (2D) system with low-electron density,  screening of Coulomb interaction via polarization is not effective.
Instead we assume that our 2D layer of Rashba material is screened by a proximate conducting gate, see Fig.~\ref{fig:setup}, which also provides a simple way to
control the (electro-)chemical potential of the system by varying the gate voltage.
The Coulomb potential, describing instantaneous screening by the gate's image charges, takes the following form:
\begin{equation}
U_{\bm q} = \frac{e^2}{2\epsilon q}(1-e^{-qd}),
\label{eq:Coulomb_gated}
\end{equation}
 where $d/2$ is separation between the gate and the layer and $\epsilon$ is dielectric constant of the material in which two-dimensional conducting layer is formed (semiconducting heterostructure).

\section{BEC regime: pair wave function approximation\label{sec:pairs}}

In this Section, we analyze the ground states of the low-density Fermionic gas described by Hamiltonians~\eqref{eq:H0}, \eqref{eq:attraction} and \eqref{eq:Coulomb_Ham}. Specifically, we study the evolution of the ground state of the system as the gate voltage is adjusted to control the position of the electrochemical potential with respect to the bottom of the moat band, assuming it remains below the moat band bottom, $\Lambda \geq 0$ in Fig.~\ref{fig:bands}: we refer to this situation as the BEC regime, see below.

For non-interacting particles, $\Lambda \geq 0$ would mean that particles do not enter the two-dimensional layer, which remains insulating. However, in the presence of particle attraction (the Cooper channel), there can be bound states of pairs of particles. For strong enough attraction, their energies can lie below the chemical potential for pairs (twice the chemical potential for single particles). In that case, pre-formed pairs would actually enter the system, and form a Bose-Einstein condensate (BEC) of pairs. We refer to this kind of situation as the BEC limit.

It is clear that since there is a finite binding energy required for a pair to enter the system, the spatial extent of such a pair is also finite, being determined by roughly the inverse binding energy. On the other hand, for the range of parameters where the pairs are just (energetically) able to enter the system, their density is vanishingly small, and one indeed deals with the case of a dilute Bose gas of pairs.

The diluteness of the pair gas implies that in order to find the conditions at which the BEC occurrs, one only needs to determine the conditions required for the first pair to enter the system. Therefore, in this Section we solve the single-pair Schr\"odinger equation to determine when the pair bound state goes under the chemical potential for pairs. Since the chemical potential plays only an auxiliary role in such considerations, setting the value below which the bound state energy needs to go, one can formally solve the single-pair problem even for $\Lambda<0$. This is especially relevant in the case with Coulomb repulsion, which effectively increases the pair fugacity above $2\Lambda$, hence $\Lambda=0$ is not special within the single-pair problem in that case. However, the physical starting point  - a single pair in the system - is not correct in the $\Lambda<0$ case, since one must start with a Fermi sea, hence we do not consider it.

\subsection{Neutral Fermions with short-range attraction \label{sec:pairs_no_Coulomb}}

We start with the case without the Coulomb interaction, pertaining to neutral Fermions with short-range attraction. In order to model and study this limit we introduce a simple ansatz wave-function made of a superposition of Cooper pairs at different possible momenta, which is an exact eigenstate of the pairing Hamiltonian:
\begin{equation}
|\Psi\rangle = \sum_{\bm p} a_{\bm p} c_{\bm p}^\dagger c_{{\bm {-p}}}^\dagger |0\rangle .
\label{eq:psi_0}
\end{equation}
The form of this wave function is dictated by the form of the pairing Hamiltonian, Eq.~\eqref{eq:attraction}, and in the absence of the Coulomb repulsion (when $H = H_0 + H_p$) the wave function~\eqref{eq:psi_0} is an exact eigenstate of the total Hamiltonian. The Schr\"odinger equation $H|\Psi\rangle = E|\Psi\rangle$ gives us the following form for the bound state wave-function in the momentum-space
\begin{equation}
\alpha_{\bm p} = \frac{1}{2\xi_p -E}\frac{g}{V}\sum_{\bm k}\alpha_{\bm k} \equiv
\frac{c}{2\xi_p- E},
\label{eq:bound-wavefunction}
\end{equation}
where $\alpha_{\bm p} = a_{\bm p} e^{-i\theta_{\bm p}}$, $\xi_p$ is given by Eq.~\eqref{eq:xi}, and $c$ is a constant given by the integral over ${\bm k}$. Solving Eq.~\eqref{eq:bound-wavefunction} self-consistently leads to a simple integral equation for $E$ as follows:
\begin{equation}
1 = \frac{g}{V} \sum _{\bm p} \frac{1}{2\xi_p -E},
\label{eq:eigeq-pairs-no coulomb}
\end{equation}
Approximating the integral is straightforward and gives us
\begin{equation}
E = 2\Lambda - \frac{g^2p_0^2}{8\kappa}.
\label{eq:bound energy}
\end{equation}
A negative value of $E$ implies that it is energetically favorable for a pair to enter the system. Therefore, the condition $E=0$ determines the critical value of the coupling constant, $g$, at which a condensate forms in the system (at a given value of $\Lambda$). It yields the critical interaction strength for the superconductor-insulator transition in the system of neutral fermions, $g_{c0}$
\begin{equation}
g_{c0} = \frac{4\sqrt{\kappa \Lambda}}{p_0}.
\label{eq:g_c0}
\end{equation}
 This results coincides with that of Ref.~\onlinecite{Chamon}, which was obtained using field-theoretical methods. Here, we just emphasize the very simple nature of this problem, which reduces to calculating a pair bound state.

Using the treatment presented above, one can also easily confirm that at the critical value of $g$, when the boundstate goes under the chemical potential for pairs, the corresponding wave-function has finite spatial extension --  the pair size. It is easy to check that this characteristic length scale is given by
\begin{equation}
\zeta = \sqrt{\frac{\kappa}{\Lambda}}.
\label{eq:pair size}
\end{equation}

\subsection{Effects of Coulomb Repulsion \label{sec:pairs_Coulomb}}

Now if we include the Coulomb repulsion, $H_c$, in the Hamiltonian, the Schr\"odinger equation becomes
\begin{equation}
(2\xi_q - E)\alpha_{\bm q} + \frac{1}{V}\sum_{\bm k} \tilde{U}_{\bm {q,k} }\alpha_{\bm k} =
\frac{g}{V}\sum_{\bm p} \alpha_{\bm p},
\label{eq:Sch_pair_Coulomb}
\end{equation}
where $\tilde{U}_{{\bm q,\bm k}} = U_{{\bm q-\bm k}}(1+\cos(\theta_{\bm q} - \theta_{\bm k}))/2$.

The key difference brought about by the Coulomb repulsion is the fact that now the bound state formation requires a finite strength of the attraction. In the case of pure attraction, the bound state exists at any positive value of $g$, see \eqref{eq:eigeq-pairs-no coulomb}, while a finite critical value $g_{c0}$ \eqref{eq:g_c0} follows from the condition that the bound state must be $2\Lambda$ deep in order for pairs to enter the system. Below, we will refer to the critical value of $g$ needed for the bound state formation as $g_{cb}$, keeping the notation $g_c$ for the value of $g$ that marks the onset of superconductivity.

The form of solutions to this equation depends on the strength of screening. One should distinguish the limits of strong, $d p_0\ll1$, and weak, $d p_0\gg1$, screening, which we discuss in detail below.

\subsubsection{Strong screening}

When the screening is strong, i.e. when $dp_0 \ll 1$, $U_q$ (defined in \eqref{eq:Coulomb_gated}) can be approximated by the constant $e^2 d /2\epsilon $, reflecting the fact that it becomes a short-range contact-like interaction. Therefore, in that case the Coulomb term in \eqref{eq:Sch_pair_Coulomb} can be absorbed in the constant attraction term on the right hand side. Absorbing the Coulomb term, one should note that due to the presence of the angular factor, $(1+\cos(\theta_{\bm q} - \theta_{\bm k}))/2$, it would diminish by a factor of $1/2$ over the angular integral. With this consideration we obtain a renormalized effective attraction
\begin{equation}
g_{\rm eff} = g - \frac{e^2 d}{4\epsilon}.
\label{eq:geff-pair-str-scr}
\end{equation}
It is obvious that the value of $g$ required for the formation of a bound state is
\begin{align}\label{eq:gcbsmalld}
    g_{cb}= \frac{e^2 d}{4\epsilon}.
\end{align}
In turn, just like in the case of pure attraction, the onset of superconductivity is marked by the value of $g$ required for the bound state to cross the value of the chemical potential for pairs, $2\Lambda$. As a result, strong screening of the Coulomb interaction by the conducting gate results in a weak enhancement of the critical strength of attraction $g_c$,
\begin{equation}
g_c = g_{c0} + \frac{E_c d}{2p_0},
\label{eq:2part_gc_coulomb_str_scr}
\end{equation}
where $g_{c0}$ is given by Eq.~\eqref{eq:g_c0}, and we defined $E_c = e^2 p_0 / 2\epsilon$ as the characteristic Coulomb energy scale in the moat band.

\begin{figure}
\includegraphics[width=\columnwidth]{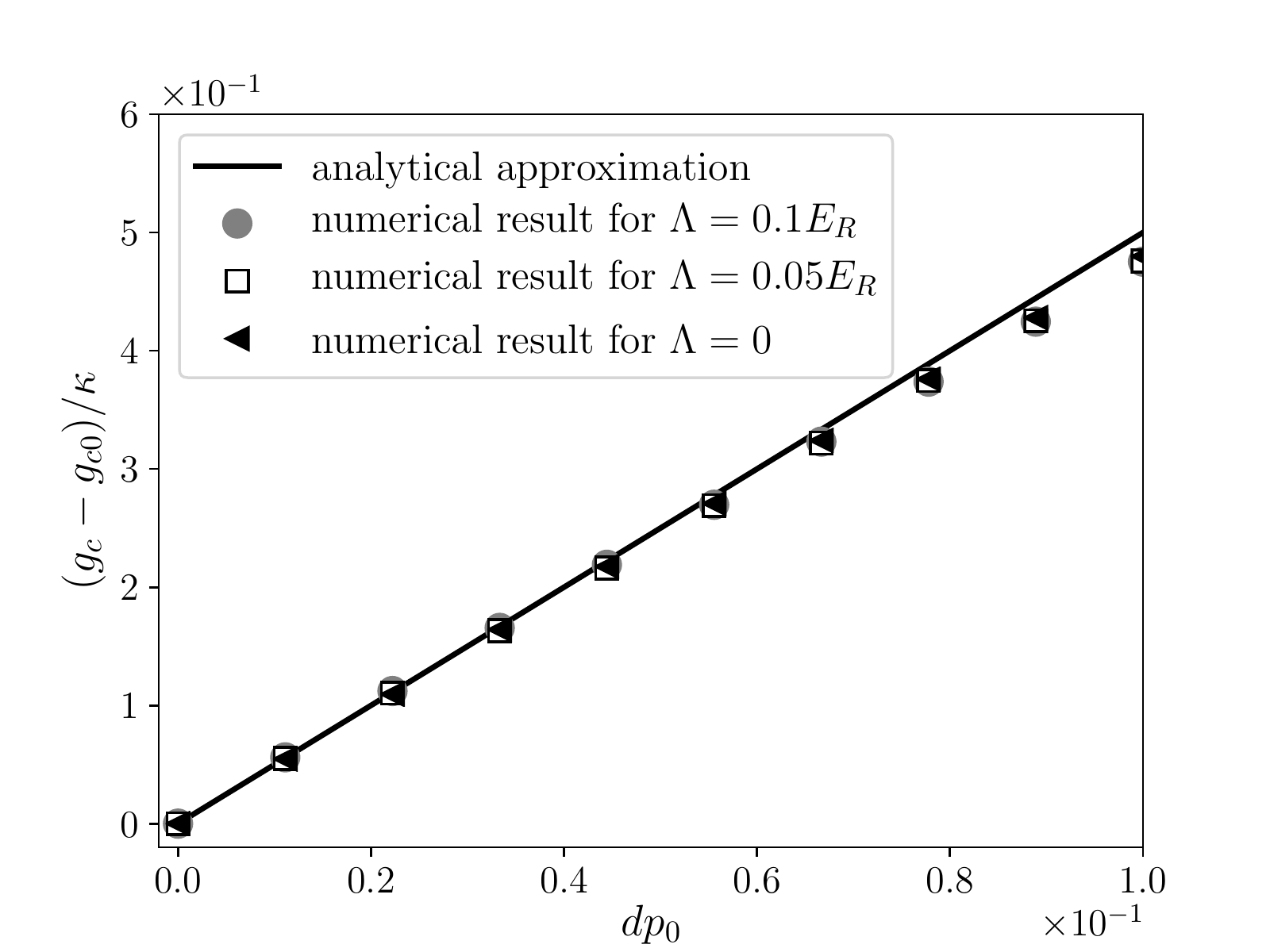}
\caption{Comparing numerical and analytic $d$-dependence of Coulomb correction to $g_c$ (Eq.\eqref{eq:2part_gc_coulomb_str_scr}) for $E_c = 10E_R$ and for three different $\Lambda$'s in the range $dp_0 \ll 1$.}
\label{fig:gc_d}
\end{figure}

In Fig.~\ref{fig:gc_d} the analytic approximation \eqref{eq:2part_gc_coulomb_str_scr} is compared with the results found by solving the Schr\"odinger equation \eqref{eq:Sch_pair_Coulomb} numerically. One can see that the strong-screening results - the linear in $d$ enhancement of the critical attraction strength - persists to $d p_0\sim 0.1$.

\subsubsection{Weak screening}\label{sec:SPWFweakscreening}
In the remainder of this Section, we focus on determining the condition for the appearance of the bound state in the two-particle spectrum in the limit of weak screening, $dp_0\gg1$, for which the Coulomb repulsion of Eq.~\eqref{eq:Coulomb_gated} is strong and $q$-dependent. We will also focus on describing qualitative features of this bound state. Full numerical solution of Eq.~\eqref{eq:Sch_pair_Coulomb} is deferred till Section~\ref{sec:BCS}, where it will be used to determine the condensation line on the BEC side of the phase diagram in the $(g,\Lambda)$ plane.

 While the Schr\"odinger equation~\eqref{eq:Sch_pair_Coulomb} can indeed be solved numerically, considerable insight into the the problem can be gained by transforming it into an effective Schr\"odinger equation in auxiliary one-dimensional ``real'' space, which is reciprocal to the radial direction in the momentum space.

The transformation proceeds by reducing Eq.~\eqref{eq:Sch_pair_Coulomb} to a radial equation (for a rotationally-symmetric $\alpha_{\bm q}=\alpha_q$):
\begin{align}
2\xi_q \alpha_{ q} -gp_0\int_p\alpha_{p}+ \int_{k} {\cal{K}}(q,k)\alpha_{ k} = E\alpha_{q},
\label{eq:1DSch_pair}
\end{align}
where the Coulomb kernel ${\cal K}(q,k)$ is defined as
\begin{equation}
{\cal{K}}(q, k) = \frac{E_c}{2\pi}\int_0^{2\pi}\frac{\cos^2(\theta_{\bm{qk}}/2)(1-e^{-d|{\bm {q-k}}|})}
{|{\bm {q-k}}|}d\theta,
\label{eq:K-function}
\end{equation}
$\theta_{\bm {qk}}$ being the angle between ${\bm q}$ and ${\bm k}$, and we introduced a shorthand notation
\begin{align}
  \int_p\equiv \int^{\infty}_{-\infty}\frac{dp}{2\pi}.
\end{align}

Furthermore, as explained above, we are interested in finding two-electron bound states with vanishing energy, $E\to 0$, for the purpose of finding the boundary of the normal-superconducting transition. For $\Lambda\ll E_R$, only momenta in the close vicinity of $p_0$ participate in the formation of such shallow bound states, which is enforced by the term with $\xi_q$ in Eq.~\eqref{eq:1DSch_pair}. Combined with the fact that the Coulomb kernel $\cal{K}$ is sharply peaked around $q=k$, this allows us to count all momenta from $p_0$ and extend all integrations over the magnitude of such reduced momenta to infinite limits, as well as substitute
\begin{align}\label{eq:ReducedCoulombKernel}
  {\cal{K}}(q,k)\to {\cal{K}}\left(p_0+\frac{q-k}{2},p_0-\frac{q-k}{2}\right)\equiv {\cal{K}}(q-k).
\end{align}
This implies that Eq.~\eqref{eq:1DSch_pair} maps onto a 1D Schr\"odinger equation, in which $2\xi_q=2\kappa (q-p_0)^2+2\Lambda\to -2\kappa \partial_x^2+2\Lambda$ plays the role of an effective 1D kinetic energy (shifted by $2\Lambda$, as we count the energeis from the chemical potential); $-gp_0\delta(x)$ represents the short-range attractive potential; ${\cal K}(x)$ - the Fourier transform of ${\cal K}(q)$, is a long-range repulsive potential.

As demonstrated in Appendix~\ref{appendix:potential}, for relatively small $q\ll p_0$, which participate in the pairing problem at realistic values of parameters, one can approximate ${\cal K}(q)$ with
\begin{align}\label{eq:CoulombKsmallq}
    {\cal{K}}(q)\approx \frac{E_c}{2\pi p_0}\ln\frac{4\beta^2 d^2p_0^2}{1+\frac{\beta^2}{4}d^2q^2},
\end{align}
where $\beta=e^{\gamma}\approx 1.78$, with $\gamma\approx 0.58$ being the Euler-Mascheroni constant. Expression~\eqref{eq:CoulombKsmallq} is of interpolation nature, and connects the $q\to 0$ limit of ${\cal K}(q)$ with the $q\gg 1/d$ limit.

The ``real''-space potential that corresponds to the kernel~\eqref{eq:ReducedCoulombKernel} is derived in Appendix~\ref{appendix:potential}, and is given by Eq.~\eqref{eq:potential_appendix}, which we repeat here for convenience:
\begin{align}\label{eq:potential}
    U(x)=\frac{E_c}{2\pi p_0 |x|}\exp\left(-\frac{2|x|}{\beta d}\right).
\end{align}
This potential defines the behavior of the effective 1D quantum-mechanical problem down to distances $|x|\gtrsim 1/p_0$. At smaller scales, the large-momentum behavior of ${\cal{K}}(q,p)$ becomes important, and Eq.~\eqref{eq:ReducedCoulombKernel} ceases to hold.

From now and until the end of this Section, we will measure all energies in units of $E_R$, all momenta in units of $p_0$, all coordinates in units of $1/p_0$, and $g$ in units of $\kappa$. Then the effective Schr\"odinger equation in real space is written as
\begin{align}\label{eq:effectiveSE}
    -2\alpha''(x)-g\delta(x)\alpha(x)+\frac{E_c}{2\pi |x|}e^{-\frac{2|x|}{\beta d}}\alpha(x)=(E-2\Lambda)\alpha(x).
\end{align}
To qualitatively understand the solutions of Eq.~\eqref{eq:effectiveSE}, we start with the $d\to\infty$ limit when $e^{-\frac{2|x|}{\beta d}} \to 1$. First, we focus on finding the critical value of $g$, $g_{cb}$, at which the first bound state emerges below the continuum -- this signals the appearance of a two-particle bound state in the original problem. To find $g_{cb}$, we set $E=2\Lambda$, and retain only the solution that does not grow indefinitely for $x\to\infty$. At distances $|x|\gtrsim 1$, such solution is
\begin{align}\label{eq:SEsolution}
    \alpha_0(x)=A \sqrt{|x|} K_1\left(\sqrt{\frac{E_c |x|}{\pi}}\right),
\end{align}
where $A$ is an arbitrary constant, and $K$ is the modified Bessel function of the second kind. Note that the asymptotic behavior of this function at large values of the argument is given by
\begin{align}
    K_1(z)\sim \sqrt{\frac{\pi}{2z}}e^{-z}.
\end{align}
This implies that even though the state $\alpha_0$ has energy infinitesimally below the continuum, it has a finite spatial extent $\alpha(x) \sim \exp(-\sqrt{|x|/\zeta_0})$, determined by
\begin{align}
    \zeta_0=\frac{\pi}{E_c}\gg1.
    \label{eq:zeta0}
\end{align}
(Note that in dimensionful units $\zeta_0 = \pi E_R/(E_c p_0)$.)
The critical value of $g$ is found from the condition that $\alpha_0(x)$ has a jump in derivative at the origin determined by $g$ -- the strength of the attractive $\delta$-function -- in the standard way. We note that for $x\to 0$, the solution presented in Eq.~\eqref{eq:SEsolution} has a logarithmically divergent derivative. This divergence is unphysical, since the form of the Coulomb potential is not valid near the origin. Therefore, we cut-off the divergence at $x\sim 1$, to obtain
\begin{align}\label{eq:criticalglarged}
g_{cb}\approx\frac{4|\alpha_0'(x\sim1)|}{\alpha_0(0)}=\frac{E_c}{\pi}\ln\frac{4\pi}{\beta^2 E_c}.
\end{align}
It is clear that the critical $g_{cb}$ given by Eq.~\eqref{eq:criticalglarged} also persists for finite $d\gtrsim \zeta_0$.

In turn, for $1\ll d\lesssim \zeta_0$, the repulsive potential  behaves qualitatively like a $\delta$-functional one, simply renormalizing the strength of the attractive potential. That is,
\begin{align}
    U(x)\approx 2\int^{d}_{1}dx' \frac{E_c}{2\pi x'}\exp\left(-\frac{2x'}{\beta d}\right)\,\delta(x).
\end{align}
Performing the integral, and taking the $d\gg1$ asymptotics of the result, we observe that in this limit the repulsive interaction  renormalizes $g$ by
\begin{align}\label{eq:medium_screening_renormalization}
    g\to g-\frac{E_c}{\pi}\ln\frac{d}{2}.
\end{align}
The bound state appears when the renormalized strength of the $\delta$-function potential is positive, hence in this regime
\begin{align}\label{eq:eq:criticalgmediumd}
    g_{cb}=\frac{E_c}{\pi}\ln \frac{d}{2}.
\end{align}
Note that for $d\sim \zeta_0$, Eqs.~\eqref{eq:criticalglarged} and~\eqref{eq:eq:criticalgmediumd} are parametrically the same.

\begin{figure}[h]
\includegraphics[width=0.8\columnwidth]{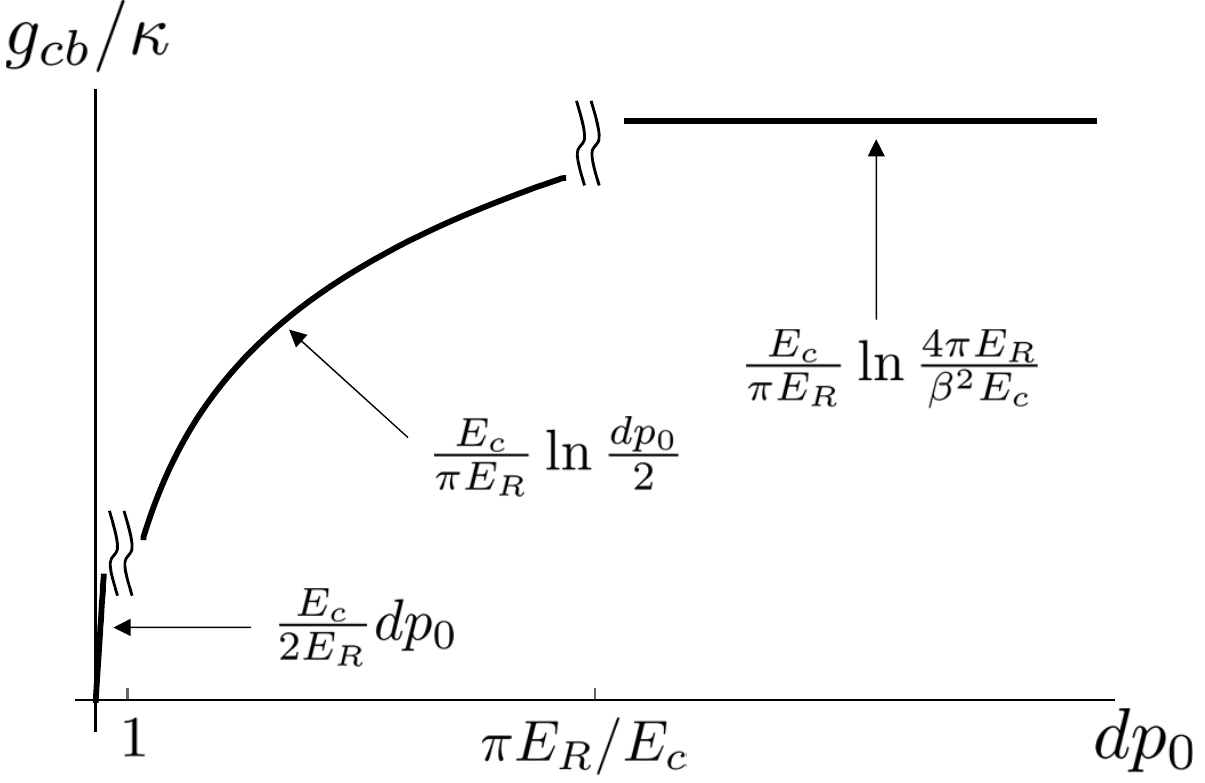}
\caption{Critical $g_{cb}$ for the two-particle bound state formation in various regimes of screening.}
\label{fig:gcb}
\end{figure}

Taken together, Eqs.~\eqref{eq:gcbsmalld},~\eqref{eq:criticalglarged}, and ~\eqref{eq:eq:criticalgmediumd} determine the dependence of the critical attraction  strength required to bind a two-particle bound state for all values of $d$. This dependence is schematically illustrated in Fig.~\ref{fig:gcb}.

The effective Schr\"odinger equation~\eqref{eq:effectiveSE} can also help us determine the dependence of the bound state energy, $E_b$, on $g>g_{cb}$. To this end, we write this equation for $\alpha_0$, as well as for $\alpha_g$, the latter being assumed to solve the equation for $g$ above the critical value, and take into account that $2\Lambda-E_b>0$ :
\begin{align}\label{eq:bindingenergyequation}
    &-2\alpha_0''(x)-g_{cb}\delta(x)\alpha_0(x)+U(x)\alpha_0(x)=0,\nonumber\\
    &-2\alpha_g''(x)-g\delta(x)\alpha_g(x)+U(x)\alpha_g(x)=-(2\Lambda-E_b)\alpha_g(x).
\end{align}
We now multiply the first equation with $\alpha_g$, and the second one with $\alpha_0$, subtract one from the other, and integrate over the entire axis. We immediately get
\begin{align}\label{eq:Eb}
    2\Lambda-E_b(g)=\frac{\alpha_0(0)\alpha_g(0)}{\int dx \alpha_0(x)\alpha_{g}(x)} (g-g_{cb}).
\end{align}
The value of the prefactor in front of $g-g_{cb}$ depends on $d$ and the value of $g$ as compared to $g_{cb}$. In two cases, either for $d\lesssim \zeta_0$ and any $g$, or for $g\gg g_{cb}$, the size of the bound state in the effective $\delta$-potential is determined by its energy, $\zeta_g\propto 1/\sqrt{\varepsilon_{cb}(g)}$. Thus we have
\begin{align}
   \frac{\alpha_0(0)\alpha_g(0)}{\int dx \alpha_0(x)\alpha_{g}(x)} \propto  \sqrt{2\Lambda-E_b(g)},
\end{align}
and hence
\begin{align}\label{eq:Ebquadratic}
    2\Lambda-E_b\propto (g-g_{cb})^2.
\end{align}.

In the limit of large $d\gtrsim \zeta_0$, and for  $g-g_{cb}\ll g_{cb}$, we can estimate
\begin{align}
\frac{\alpha_0(0)\alpha_g(0)}{\int dx \alpha_0(x)\alpha_{g}(x)} \propto  \frac{1}{\zeta_0   },
\end{align}
and conclude that
\begin{align}\label{eq:Eblinear}
    2\Lambda-E_b\propto (g-g_{cb}).
\end{align}

In principle, the boundary of the insulator-superconductor transition can be obtained by solving Eq.~\eqref{eq:Eb}, and setting $E_b(g)=0$. We will not pursue the task of its determination, deferring it until Section~\ref{sec:BCS_with-coulomb}, where many-body aspects of the condensation problem are discussed. The intuition used in that Section, however, heavily rests on the results of this one.

\section{Variational BCS wave-function}
\label{sec:BCS}

We now improve our treatment by using a BCS-type variational wave function.
We restrict ourselves to the study of the many-body ground state of the system at $T = 0$, and aim to investigate the situation with small but finite electron density in the moat band.
Later on, we use the zero-temperature results to estimate the critical temperature of the system.

It should be clarified here that by considering a BCS variational wave function, we are assuming that the normal state of the model at $\Lambda < 0$, i.e. when the chemical potential is inside the moat band,
is a Fermi liquid one. This assumption is based on previous studies \cite{berg_kivelson, berg_FM} of the phase diagram of the repulsively-interacting electrons
which found that other, more exotic, electron phases in this interesting regime - such as Wigner crystal, nematic and ferromagnetic phases - require significant Coulomb repulsion (see, for example, Fig.1 of Ref.\onlinecite{berg_FM}). Such a strong repulsion is not consistent with the superconducting state we are considering here.

As shown below, the BCS wave function allows us to study the behavior of the superconducting order parameter $\Delta$ across the crossover region between the BEC and BCS limits, in the absence and presence of Coulomb interaction.

The variational wave function that we consider is
\begin{equation}
|\Psi_{BCS}\rangle = \prod_{p_x > 0} (u_{\bm p} + e^{i\theta_{\bm p}}v_{\bm p}
 c_{\bm p}^\dagger c_{{\bm {-p}}}^\dagger)|0\rangle,
\label{eq: BCS-fun}
\end{equation}
where $u_{\bm p}$ and $v_{\bm p}$ are the variational functions; $v_{\bm p}$ is multiplied by the phase factor which is chosen to compensate for the phase factors in the attractive channel, Eq.~\eqref{eq:attraction}.
Similar to Section~\ref{sec:pairs} we first study the model without repulsion and then will investigate the effects of Coulomb repulsion.

\subsection{Phase diagram for neutral fermions}
\label{sec:BCS_no-coulomb}
\begin{figure}[h]
\includegraphics[width=\columnwidth]{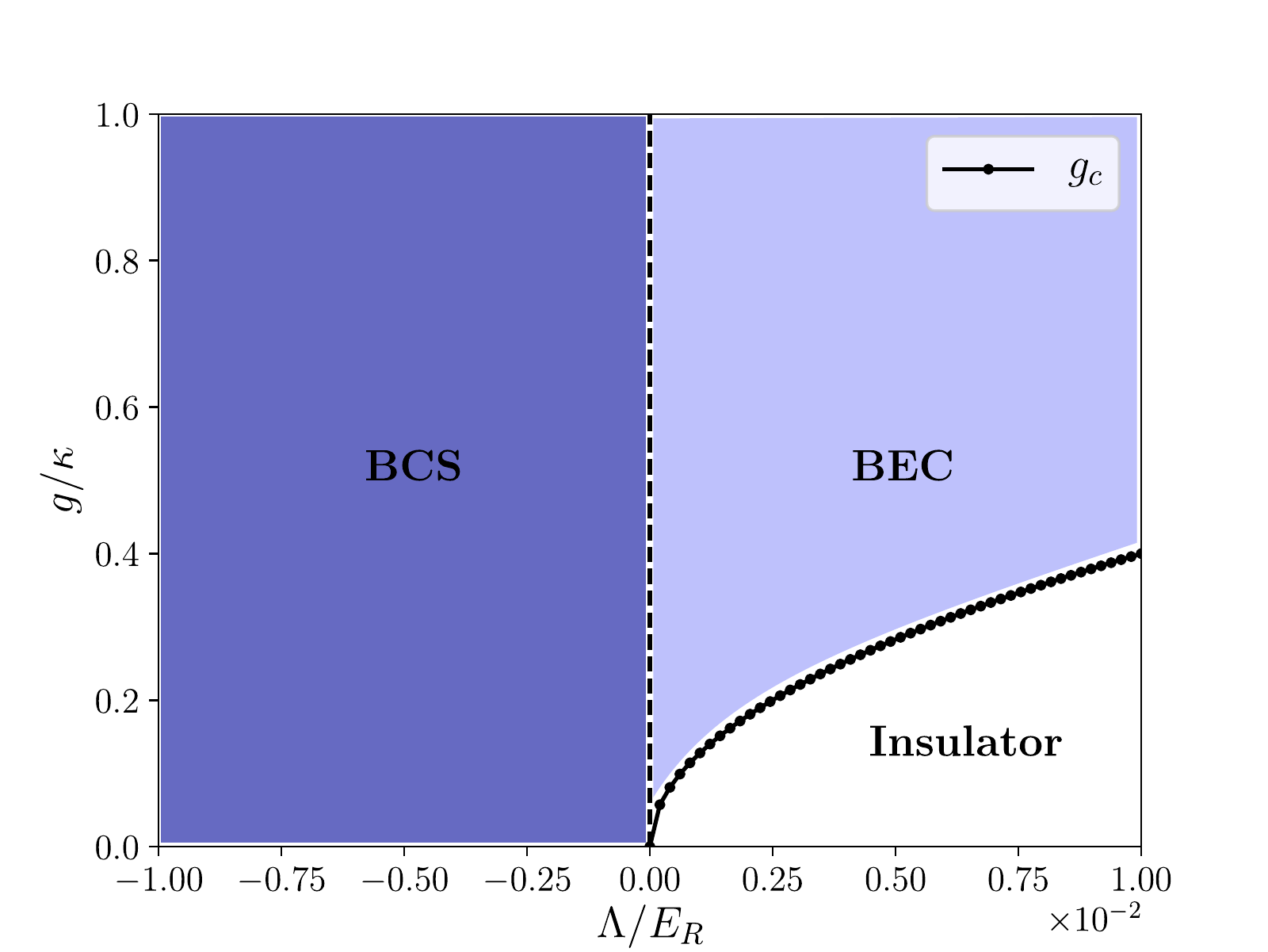}
\caption{ (Color online) Phase diagram in the absence of Coulomb repulsion. $g_c$ line on the $\Lambda$ positive side is drawn based on Eq.~\eqref{eq:g_c0}.}
\label{fig:PDneutral}
\end{figure}

We start with summarizing the phase diagram in $(g,\Lambda)$ coordinates for the case of neutral fermions with attraction. It is depicted in Fig.~\ref{fig:PDneutral}.

Qualitatively, for $g>0$, the phase diagram contains two phases: an insulator, and a superconductor. The superconducting phase is stable for all values of $\Lambda\leq0$ (chemical potential is inside, or touches the majority band), and for $g>g_c(\Lambda)$ for positive $\Lambda$ (when the chemical potential is below the majority band), see Fig.~\ref{fig:PDneutral}. For $\Lambda>0$, the transition from the insulator into a superconductor upon increasing $g$ occurs into a BEC-type superconducting phase, which is represented by a dilute gas of bound pairs of electrons. For $\Lambda\leq 0$, the superconducting phase has BCS character, in which it is no longer permissible to disregard the fermionic nature of the electrons comprising the pairs in the condensate, and think of it as a dilute Bose gas.

There are two notable lines in the phase diagram of Fig.~\ref{fig:PDneutral}: the line of superconductor-insulator transition, and the line of BEC-BCS crossover. The superconductor-insulator transition line is determined from the condition that the bound state energy for electronic pairs lies at the chemical potential for pairs (chosen to be zero in our analysis). In other words, counted from the bottom of the two-electron continuum, the binding energy of the pair must be equal to $2\Lambda$. The corresponding critical value of $g$ was calculated in Section~\ref{sec:pairs_no_Coulomb}, and is given by Eq.~\eqref{eq:g_c0}.

The BEC-BCS crossover line has the inherent vagueness in its definition, which depends on the criterion for the crossover. Here we adopt the following logic for drawing such lines: the BEC-BCS crossover is a crossover between regimes in which the Fermi statistics of pair constituents either does not, or does matter. The importance of carrier statistics is dictated by occupations of states in momentum space. So we define the crossover as the situation in which the maximum occupation of states in the momentum space is 1/2. It will become apparent below that for the case of neutral fermions, and within the model adopted in the present paper, the BEC-BCS line is then the vertical $\Lambda=0$ one - the $g$-axis.

To get further quantitative insight into the behavior of the superconducting order parameter, we study its variation along a horizontal cut of the phase diagram in Fig.~\ref{fig:PDneutral}, that is, we vary $\Lambda$ at a given value of $g$. We will show below that the superconducting gap, $\Delta(\Lambda)$, is represented by a universal curve, which we determine below.

Minimization of \eqref{eq:H0} and \eqref{eq:attraction} over the BCS wave function \eqref{eq: BCS-fun} leads to the standard self-consistent equation \cite{jishi} for superconducting order parameter $\Delta$
\begin{equation}
1 = \frac{g}{2V}\sum_{\bm p} \frac{1}{\sqrt{\xi_p^2 + \Delta^2}},
\label{eq:self-consist-no-coulomb}
\end{equation}
where $\xi_{\bm p}$ is the single electron dispersion \eqref{eq:xi}. Integrating over the angle and noting that
 the most important part of the radial integral comes from the vicinity of $p_0$, we approximate $p dp \sim p_0 dp$ and extend the $p$-integral to $\pm \infty$. This gives
\begin{equation}
1 = \frac{1}{\pi}\int_{-\infty}^\infty \frac{dx}{\sqrt{(x^2 + \tilde{\Lambda})^2 + \tilde{\Delta}^2 }},
\label{eq:universal-Del-Lam}
\end{equation}
where $\tilde{\Lambda} = \Lambda / \Lambda_{c0}$ and $\tilde{\Delta} = \Delta / \Lambda_{c0}$ are made dimensionless with the help of $\Lambda_{c0} = g^2p_0^4 / 16 E_R$.
This integral equation holds for both positive (chemical potential below the band) and negative (chemical potential inside the moat band) $\tilde{\Lambda}$.
 Note that the characteristic energy $\Lambda_{c0}$ scales up like $g^2$.
It is clear that for reasonably small $g$ $\Lambda_{c0}$ stays much smaller than the Rashba energy $E_R$. Therefore $\tilde{\Lambda}$ can be large even in the `deep moat' limit $|\Lambda| \ll E_R$.

\begin{figure}[h]
\includegraphics[width=\columnwidth]{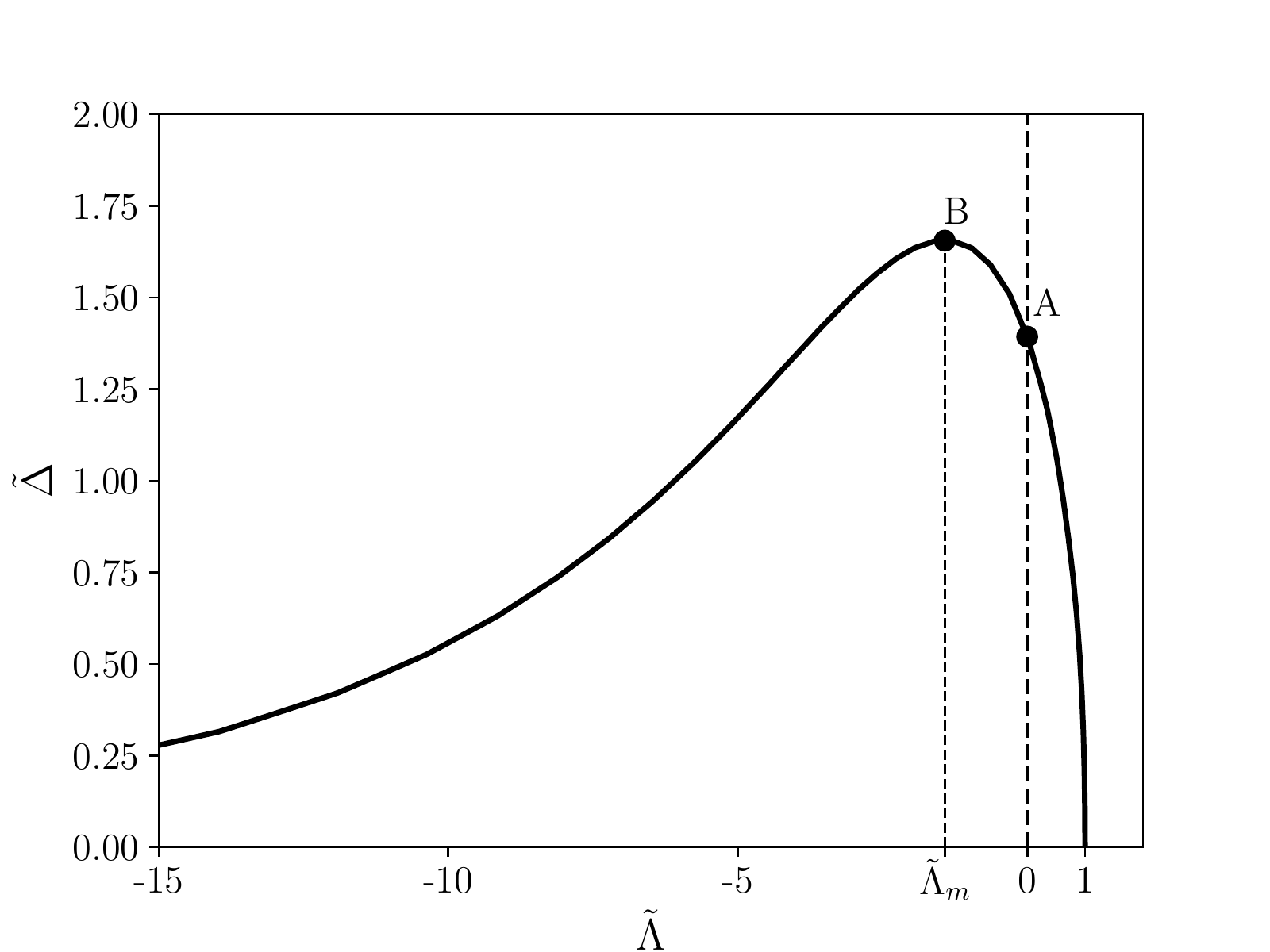}
\caption{The universal function $\tilde{\Delta}(\tilde{\Lambda})$. Two important points of the function are shown in the figure: $\bm{A}$ is the point where the chemical potential touches the band edge, which is the point of the BEC-BCS crossover, with coordinates $[\tilde{\Lambda}_0, \tilde{\Delta}_0] \approx [0,1.393]$; $\bm{B}$ is the point where the maximum of $\tilde\Delta$ is located, with coordinates $[\tilde{\Lambda}_m, \tilde{\Delta}_m] \approx [-1.424,1.655]$.}
\label{fig:Del_Lam}
\end{figure}

Fig.~\ref{fig:Del_Lam} shows numerical solution $\tilde{\Delta}(\tilde{\Lambda})$ of Eq.~\eqref{eq:universal-Del-Lam}.
Finite $\tilde{\Delta}$ for positive $\tilde{\Lambda}$ describes BEC
regime - condensed state of two-electron pairs. The order parameter vanishes at $\Lambda = \Lambda_{c0}$ with a discontinuity in its derivative.
It can be shown that asymptotic behavior of $\tilde{\Delta}$ near the transition point is $\tilde{\Delta} \sim \sqrt{\frac{8}{3}(1-\tilde{\Lambda})}$.

The other feature of function $\tilde{\Delta}(\tilde{\Lambda})$ (Fig.~\ref{fig:Del_Lam}) is that it has a maximum on the negative side,
when the chemical potential is inside the moat band.
That implies that the highest critical temperature happens when the chemical potential is slightly above the band edge.

Moreover, on the BCS side, when electrons fill up a shallow ring-shaped Fermi sea, the order parameter, $\tilde{\Delta}$, has two distinct behaviors:
when $|\Lambda|$ is of order of $\Lambda_{c0}$, the order parameter is of the same order and is proportional to  $g^2$.
For larger $|\Lambda|$, when $|\tilde{\Lambda}| \gg 1$,
the order parameter decays exponentially as $\Delta \propto |\Lambda| \exp \left(-\frac{\pi}{2}\sqrt{|\Lambda|/\Lambda_{c0}}\right)$.  To see this,
expand $(x^2 - |\Lambda|)^2 \approx 4 |\Lambda| (x-|\Lambda|^{1/2})^2$ and approximate the integral in \eqref{eq:universal-Del-Lam} by the region
around $x\approx |\Lambda|^{1/2}$.
The large-$|\Lambda|$ has a familiar form $\exp (-{\text{const}} / g)$ for the order parameter in a standard BCS theory suggesting a superconducting state with strongly overlapping electron pairs at large enough $|\Lambda|$.

Finally, the BEC-BCS crossover line can be deduced from the usual BCS coherence factor, $v_p$, whose square defines the occupation of states in the momentum space:
\begin{align}
    v^2_p=\frac12\left(1-\frac{\xi_p}{\sqrt{\xi^2_p+\Delta^2}}\right).
\end{align}
It is clear that for $\Lambda>0$, and hence $\xi_p>0$, all occupation numbers are smaller than 1/2. In turn, for $\Lambda<0$, when $\xi_p$ changes sign at the Fermi momenta, there is a region in momentum space -- where there is ``water'' (i.e. electron liquid) in the moat -- where the occupation numbers exceed 1/2. For $\Lambda=0$, regardless of $g$, the maximum occupation is exactly 1/2, and is reached at $p=p_0$, at the bottom of the moat. Hence the $\Lambda=0$ is the BEC-BCS crossover line according to the criterion we chose, as shown in Figure \ref{fig:PDneutral}.

\subsection{Phase diagram with Coulomb repulsion}
\label{sec:BCS_with-coulomb}
\begin{figure}[h]
\includegraphics[width=\columnwidth]{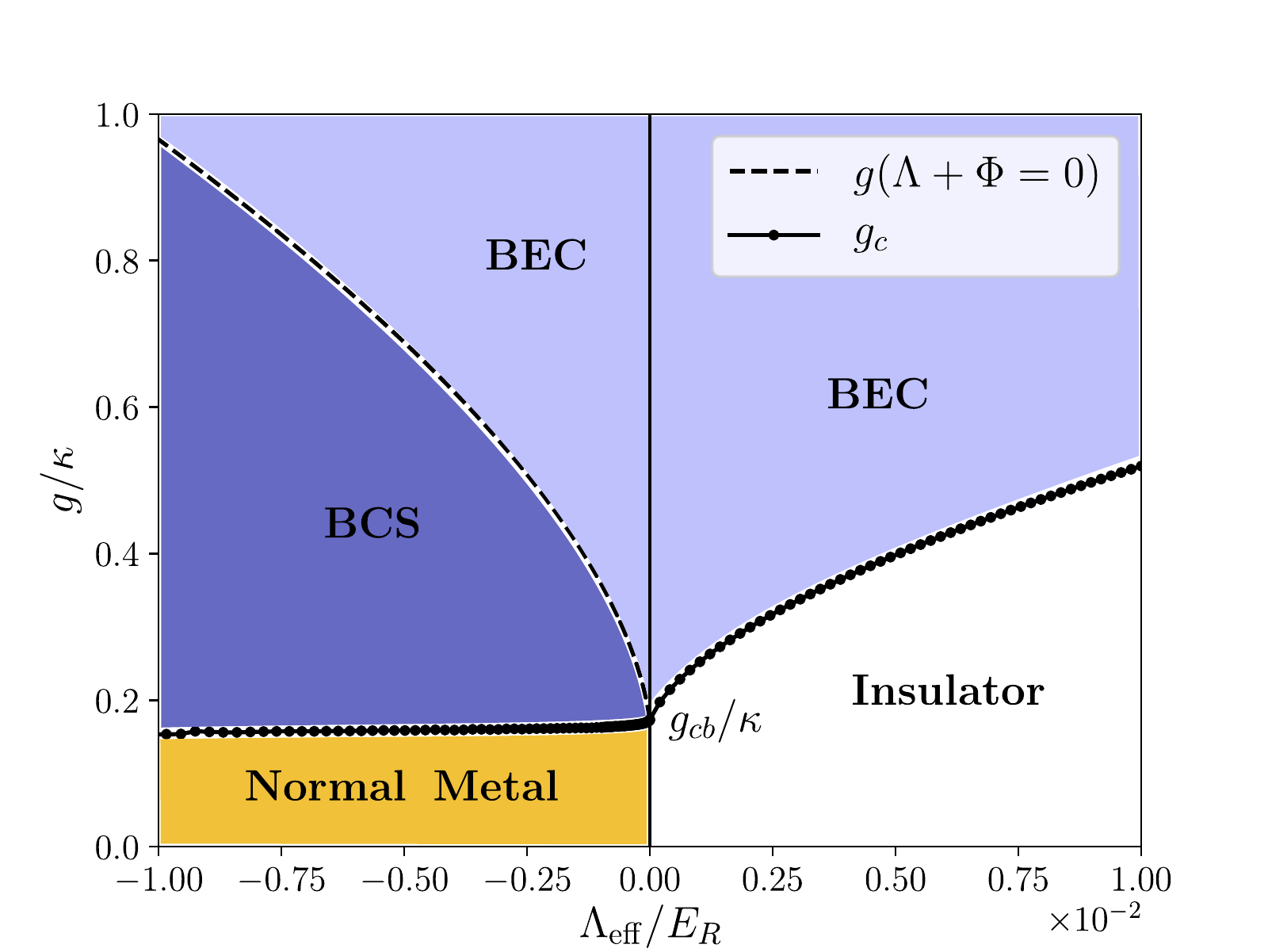}
\caption{(Color online) Phase diagram in the presence of Coulomb repulsion. The boundaries were found numerically for a weakly screened Coulomb case with parameters: $dp_0 = 1000$ and $E_c = E_R/10$. $\Lambda_{\rm eff}$ on the $x$-axis corresponds to the value of $\Lambda + \Phi$ in the normal state, which is the same as bare $\Lambda$ on the insulator side and is renormalized strongly on the metallic side according to Eq.~\eqref{eq:L+P}.}
\label{fig:PDcoulomb}
\end{figure}

Inclusion of Coulomb repulsion between electrons changes the phase diagram, which is now shown in Fig.~\ref{fig:PDcoulomb}. It is convenient to describe phases appearing in this diagram starting with the behavior of the system along $\Lambda=0$ line. As we know from Section~\ref{sec:pairs_Coulomb}, in the presence of the Coulomb repulsion, there is a finite value of $g$, which we denoted with $g_{cb}$, needed to achieve binding of pairs even at $\Lambda=0$. Therefore, for $0\leq g< g_{cb}$, the $\Lambda=0$ line is the boundary between the insulating phase on the right, and the normal metal phase on the left. In other words, the system behaves as a semi-metal with diverging one-dimensional density of states for $g<g_{cb}$, and $\Lambda=0$. To the left of this semimetallic phase, there lies a normal metal phase. The doping level of this metal is lower than the nominal one prescribed by $|\Lambda|$, due to electrostatic effects.

Superconducting phases can be reached from both metallic and insulating phases. As in the case of neutral fermions, one reaches BEC- or BCS-like states increasing $g$ on the $\Lambda\gtrless 0$, respectively. Exactly at $\Lambda=0$, one goes into a BEC superconductor, since the BEC-BCS line lies on the left of the $g$-axis in this case, see below.

Having described the qualitative features of the phase diagram for charged fermions, we elaborate now on its quantitative aspects. The Coulomb repulsion brings two new terms to the total energy, one through the Cooper channel (third term in the equation below) and the Hartree-Fock term (forth term below),
to which the attractive interaction gives no contribution,
\begin{eqnarray}
&&\mathcal{E} = \mathcal{E}_{\text{K}} + \mathcal{E}_g
+ \mathcal{E}_{c,\Delta} + \mathcal{E}_{\text{H-F}}=
\sum_{\bm{k}}v_{\bm k}^2 \xi_k -
 \frac{g}{2V}\sum_{{\bm {kk'}}}u_{\bm k} v_{\bm k} u_{\bm {k'}} v_{\bm {k'}}\nonumber\\
&&+\frac{1}{2V}\sum_{{\bm {kk'}}}u_{\bm k} v_{\bm k} u_{\bm {k'}} v_{\bm {k'}}
\tilde{U}_{\bm {k,k'}} + \frac{1}{2V}\sum_{{\bm {kk'}}}v_{\bm k}^2 v_{\bm {k'}}^2
\left(\tilde{U}_0 - \tilde{U}_{\bm {k,k'}}\right),
\label{eq:BCS energy}
\end{eqnarray}
where $\tilde{U}_{{\bm {k,k'}}} = U_{{\bm {k-k'}}}(1+\cos(\theta_{\bm k} - \theta_{\bm {k'}}))/2$, as before.
Minimizing the energy over $v_{\bm k} = \sin\alpha_{\bm k}$ (with $u_{\bm k} = \cos\alpha_{\bm k}$) and introducing fields
\begin{eqnarray}
\Delta_{\bm k} &=&  \frac{1}{V}\sum_{\bm {k'}} u_{\bm {k'}} v_{\bm {k'}} \big(g - \tilde{U}_{\bm {k,k'}}),\nonumber\\
\Phi_{\bm k} &=&  \frac{1}{V}\sum_{\bm {k'}} v_{\bm {k'}}^2 (\tilde{U}_0 - \tilde{U}_{\bm {k,k'}}),
\label{eq:d-ph}
\end{eqnarray}
we obtain desired coupled self-consistent equations for the superconducting and Hartree-Fock (HF) fields
\begin{align}
\label{eq:self-consist-coulomb}
& \Delta_{\bm p} = \frac{1}{2V}\sum_{{\bm {p'}}}\frac{\Delta_{{\bm {p'}}}
\left( g - \tilde{U}_{\bm {p,p'}}\right)}
{\sqrt{(\xi_{\bm{p'}} + \Phi_{\bm{p'}})^2 + \Delta_{\bm{p'}}^2}} \\ \nonumber
& \Phi_{\bm p} = \frac{1}{2V}\sum_{{\bm {p'}}}\left(1- \frac{\xi_{\bm {p'}} + \Phi_{\bm {p'}}}
{\sqrt{(\xi_{\bm {p'}} + \Phi_{\bm {p'}})^2 + \Delta_{\bm {p'}}^2}}\right)
\left(\tilde{U}_0 - \tilde{U}_{\bm {p,p'}} \right) .
\end{align}

It is clear from Eqs.~\eqref{eq:self-consist-coulomb} that the Coulomb repulsion has two-fold effect on the pairing problem. First, it changes the coupling constant in the equation for the order parameter $\Delta_{\bm p}$, $g\to g-\tilde{U}_{\bm {p,p'}}$, making it a coupling matrix. Physically, this describes interaction of electrons within a pair, and the effects of this interaction have been described in Section~\ref{sec:pairs_Coulomb}.

Second, the Coulomb repulsion leads to the appearance of the Hartree-Fock renormalization of the single-particle spectrum, described by $\Phi_{\bm p}$.
The first bracket in the equation for $\Phi_{\bm p}$ is just $2v_{\bm{p'}}^2$, twice the particle number at the momentum $\bm{p}'$. Moreover,
the second bracket in the same equation is always positive because Coulomb interaction \eqref{eq:Coulomb_gated} is maximum at $\bm{p} = 0$,
$\tilde{U}_0 = e^2 d/2\epsilon$. Therefore
the new collective field $\Phi_{\bm k}$ is a kind of ${\bm k}$-dependent chemical potential which provides a positive shift of the single particle spectrum $\xi_{\bm k}$.
This is the same as downward renormalization of the chemical potential and has the effect of `pushing out' electrons from the system.
The renormalization is proportional to the Coulomb repulsion experienced by a particle at momentum ${\bm k}$
due to all other particles (at momenta ${\bm k'}$).

In a usual good conductor such a renormalization is neglected due to the strong screening of Coulomb interaction.
This point is made explicit in classic papers on strongly coupled superconductivity, see for example \cite{Scalapino1966,Schrieffer1999} (note that $\Phi$ is denoted as $\chi$ there).
The reason is that renormalization of the spectrum, which represents a ``simple scale change'' (Ref.\onlinecite{Scalapino1966}) of electron dispersion,
is a small effect on the scale of Fermi energy $E_F$ which is present in both normal and superconducting states.
In our case, the normal state is either a band insulator, or a metal with a very small carrier density, hence the scale change due to $\Phi$ is crucial in determining whether or not electrons can populate the bottom of the moat band, either in the form of interacting electron gas or bound two-electron molecules.

Therefore the combination of low-density and weak screening (which is determined by the distance due to the external gate)
results in the HF field $\Phi_{\bm p}$ playing an unusually important role. The basic physics is that of an electrostatic
repulsion between two electrons attempting to form a bound state. In the absence of screening and retardation (our attractive potential $g$ is frequency
independent constant) the electrostatics of the pair becomes crucial. As discussion in Sec.\ref{sec:BSC_no_screen} below shows, weakly screened Coulomb interaction
strongly renormalizes bare chemical potential (see Eq.\eqref{eq:L+P} below) and squeezes electrons out, providing a kind of `Coulomb blockade' phenomenon in a
macroscopic setting. This crucial physics is missing in Ref.\onlinecite{Chamon}.

It is clear from Eq.~\eqref{eq:self-consist-coulomb} that the presence of the Coulomb interaction leads to $\bm p$-dependence of both $\Delta_{\bm{p}}$ and $\Phi_{\bm{p}}$.
However, it is not so clear if the possible solutions are radially symmetric or not. We have carefully investigated this question numerically and concluded that
the lowest energy solution of the coupled equations \eqref{eq:self-consist-coulomb} is s-wave symmetric.
We would like to add here that frequently evoked Kohn-Luttinger physics \cite{Vafek2011},
which generally promotes pairing in the finite angular momentum channel and is mediated by repulsive inter-electron interactions via Friedel-like polarization
of the fermionic background \cite{Kagan2015}, is not expected to contribute in the present situation
of a very dilute two-dimensional electron gas. In this case, one needs to include diagrams up to the third-order in repulsive interaction \cite{Chubukov1993}.
This leads to an exponentially small in the density $n$ dependence of the resulting critical temperature, 
$T_c \sim \exp[-1/n^3]$, which is not competitive with the estimate of the critical temperature based on our consideration, see
discussion at the end of this Section.

Therefore below we investigate solution with s-wave symmetry and perform angular integration to obtain one dimensional integral equations as follows:
\begin{eqnarray}
\Delta_p &=& \frac{1}{4\pi}\int_0^{p'_{\rm max}}  dp'\frac{p'\Delta_{p'}}
{\sqrt{(\xi_{p'}+\Phi_{p'})^2 + \Delta_{p'}^2}}\times\nonumber\\
&&\left(g -{\cal K}(p,p')\right),\nonumber\\
\Phi_p &=& \frac{1}{4\pi p_0}\int_0^{p'_{\rm max}} dp' p'
\left(1- \frac{\xi_{p'}+\Phi_{p'}}{\sqrt{(\xi_{p'}+\Phi_{p'})^2 + \Delta_{p'}^2}}\right)\times \nonumber\\
&&\left(E_c d - {\cal K}(p,p')\right),\nonumber\\
\label{eq:self-consis-colmb-1d}
\end{eqnarray}
where the Coulomb kernel ${\cal K}(p,p')$ is given by Eq.~\eqref{eq:K-function}. The upper momentum cut-off, $p_{\rm max}$, is taken to be $2p_0$ in numerical solutions of the above equations.

Armed with Eqs.~\eqref{eq:self-consis-colmb-1d}, we are going to consider the limits of strong ($dp_0 \ll 1$) and weak-screening ($dp_0 \gg 1$), similar to Section~\ref{sec:pairs}.

\subsubsection{strong screening: $dp_0 \ll 1$ \label{sec:BSC_str_screen}}

We start with the limit of strongly screened Coulomb interaction. In this limit
\begin{equation}
{\cal K}(p,p')\approx\frac{E_c d}{2p_0}.
\label{eq:C-approx for str screen}
\end{equation}
With this approximation, the $p$-dependence of $\Delta$ and $\Phi$ disappears,
and the effect of Coulomb interaction is to reduce $g$ by $E_c d/2p_0$ (compare with \eqref{eq:2part_gc_coulomb_str_scr}) resulting in an effective attraction
\begin{equation}
g_{\text{eff}} = g - \frac{E_c d}{2p_0}.
\label{eq:g_eff-BCS-str-screen}
\end{equation}
Following the notation used in Section ~\ref{sec:BCS_no-coulomb}, we define $\Lambda_{c,\text{eff}} = g_{\text{eff}}^2p_0^4 / 16 E_R$,
introduce dimensionless combinations $\tilde{\Lambda} = \Lambda/\Lambda_{c,{\rm eff}}$ and $\tilde{\Phi} = \Phi/\Lambda_{c,{\rm eff}}$
and extend integration limits due to the fast convergence of the integral.
The first equation in \eqref{eq:self-consis-colmb-1d} becomes
\begin{equation}
1 = \frac{1}{\pi}\int_{-\infty}^\infty \frac{dx}{\sqrt{(x^2 +
\tilde{\Lambda} + \tilde{\Phi})^2 + \tilde{\Delta}^2 }},
\label{eq:del_eq_str_scr}
\end{equation}
which is just the same as Eq.~\eqref{eq:universal-Del-Lam} upon switching $\tilde{\Lambda} + \tilde{\Phi}$ to $\tilde{\Lambda}$.
The second equation in Eq.~\eqref{eq:self-consis-colmb-1d} can be written as
\begin{equation}
\tilde{\Phi} = \frac{g - g_{\rm eff}}{\pi g_{\rm eff}}
\int_{-\infty}^\infty \left[ 1-\frac{x^2 + \tilde{\Lambda} + \tilde{\Phi}}
{\sqrt{(x^2 + \tilde{\Lambda} + \tilde{\Phi})^2 + \tilde{\Delta}^2 }} \right] dx.
\label{eq:phi_eq_str_scr}
\end{equation}

Eq.\eqref{eq:del_eq_str_scr} shows that $\tilde{\Delta}$ in terms of $\tilde{\Lambda} + \tilde{\Phi}$ is given by the same universal function that is
plotted in Fig.~\ref{fig:Del_Lam}.
Therefore, with appropriate rescaling $\tilde{\Lambda} \to \tilde{\Lambda} + \tilde{\Phi}$ one can find dependence of $\Delta$ on the bare $\Lambda$.
For instance, from Fig.~\ref{fig:Del_Lam} we know that  the order parameter $\tilde{\Delta}$ turns to zero at the critical point $\tilde{\Lambda} + \tilde{\Phi}=1$. Eq.~\eqref{eq:phi_eq_str_scr} shows
that at this point $\tilde{\Phi}$ is also zero, and therefore we conclude that $\tilde{\Lambda}=1$, that is $\Lambda=\Lambda_{c,\text{eff}}$, is the bare critical $\Lambda$ in this case too.

The same logic shows that the maximum value of $\Delta$ does not depend on the renormalization of $\Lambda$ and is given by $\Delta_{\rm m} \approx 1.65 \Lambda_{c,\text{eff}}$.

Finally, similarly to Sec.\ref{sec:BCS_no-coulomb} there too is a region where $|\tilde{\Lambda} + \tilde{\Phi}| \gg \Lambda_{c,\text{eff}}$ and the order parameter is exponentially small.
If $g/g_{\text{eff}} -1$ is very small the asymptote in terms of bare $\Lambda$ has the same form as in the case with no Coulomb interaction, because then $\tilde{\Lambda} + \tilde{\Phi} \sim \tilde{\Lambda}$.
In the other limit, when $g_{\text{eff}} \ll g$, although we get a different asymptote as far as $\Lambda$-dependence is concerned, we still get a similar $g$-dependence in
$\Delta$ that goes as $\Delta \sim \exp(-\text{const}/g_{\text{eff}})$.

To summarize, the case of strongly screened Coulomb interaction is very similar to that with no  Coulomb interaction. $\Lambda + \Phi$ plays the role of {\em effective} $\Lambda$ where, thanks to the strong screening, $\Phi$ slightly modifies the bare $\Lambda$.

\subsubsection{Weak screening: $dp_0 \gg 1$}
\label{sec:BSC_no_screen}

Now we consider the weak screening limit where distance to the screening gate $d$ is large in comparison with $1/p_0$, Coulomb interaction is only weakly screened and therefore repulsion between electrons is strong.

From the results of Section~\eqref{sec:pairs_Coulomb}, it is apparent that even in the $dp_0\gg1$ case one should distinguish two regimes: $d\lesssim\zeta_0\equiv\pi E_R/E_cp_0$, and $d\gtrsim\zeta_0$. As we saw in Section~\eqref{sec:pairs_Coulomb}, $\zeta_0$ plays the role of the size of the two-particle bound state on its appearance for $d\to \infty$. In the $d\lesssim\zeta_0$ regime, which can be called the regime of intermediate screening, one can still introduce an effective local coupling constant, see Eq.~\eqref{eq:medium_screening_renormalization}. While the electrostatic effects, described by $\Phi_{\bm p}$, are strong in this case, the basic physics is very similar to the strong screening case. Therefore, in what follows we concentrate on the $d\gtrsim\zeta_0$ limit of really weak screening.

We start the analysis of Eqs~\eqref{eq:self-consis-colmb-1d} with noting that the $p$-dependence of the $\Phi$~field can be neglected. To see that, note that for large $d p_0$ the maximum of function ${\cal K}(p,p')$ goes like $\ln(d)$, see Eq.~\eqref{eq:apppendixK0} of Appendix~\ref{appendix:potential}. Therefore the $\cal K$-term in the second bracket in the equation for $\Phi$ \eqref{eq:self-consis-colmb-1d} can be dropped in
comparison with the first, $d$ term, and the $\Phi$-equation simplifies to
\begin{equation}
\Phi = \frac{E_c d}{p_0}n(\Delta, \Phi, \Lambda),
\label{eq:Phi_eq}
\end{equation}
where the particle density $n$ is given by
\begin{equation}
n = \frac{p_0}{4\pi}\int_{-\infty}^\infty
\left[1- \frac{\xi_p + \Phi}{\sqrt{(\xi_p + \Phi)^2 + \Delta_p^2}}\right]dp.
\label{eq:density}
\end{equation}

Next, let us assume that $\Delta_p \ll \xi_p + \Phi$. We are going to show that this assumption holds as long as the pair condensate is dilute. Under this assumption, the first of Eqs.~\eqref{eq:self-consist-coulomb} can be linearized as
\begin{equation}
\Delta_{\bm p} = \frac{1}{2V}\sum_{{\bm {p'}}}\frac{\Delta_{{\bm {p'}}}
\left( g - \tilde{U}_{\bm {p,p'}}\right)}
{\xi_{\bm{p'}} + \Phi}.
\label{eq:Del-linear}
\end{equation}
We then notice that the above equation is equivalent to the single pair Schr\"odinger equation~\eqref{eq:Sch_pair_Coulomb} upon substituting the following ansatz for $\Delta_p$:
\begin{equation}
\Delta_{\bm p} = \delta (\xi_{\bm p} + \Phi)\alpha_{\bm p},
\label{eq:the-ansatz}
\end{equation}
where $\delta$ is a normalization constant to be found from Eq.~\eqref{eq:Phi_eq}. The proposed above ansatz works if one identifies
\begin{equation}
\Phi = - \frac{E_b}{2}.
\label{eq:Phi-E}
\end{equation}
The condition~\eqref{eq:Phi-E} requires $E_b < 0$, since $\Phi$ always has a positive value proportional to the density (see Eq.~\eqref{eq:Phi_eq}). In other words, we can say that if the single pair Scr\"odinger equation~\eqref{eq:Sch_pair_Coulomb} has a solution with $E < 0$, then through \eqref{eq:the-ansatz} and \eqref{eq:Phi-E} one can construct a solution for the linearized BCS equation~\eqref{eq:Del-linear} out of it. Note that for the case of $\Lambda > 0$, such a solution for \eqref{eq:Sch_pair_Coulomb} is automatically  a bound state. That is because for $\Lambda > 0$, the bottom of the continuum is always above the chemical potential. We therefore conclude that $\Phi=-E_b/2$, where, as before, $E_b$ is the two-particle bound state energy counted from the chemical potential for pairs. Since on the $\Lambda>0$ side $\Phi\neq 0$ corresponds to the presence of a superconducting condensate, we conclude that
\begin{align}
    E_b(g,\Lambda)=0
\end{align}
is the equation for the superconductor-insulator transition line, $g_c(\Lambda)$. Therefore, it can be obtained from the solution of the two-particle problem, described in Section~\ref{sec:pairs_Coulomb}. The result of numerical solution of this problem is shown in Fig.~\ref{fig:PDcoulomb}, on the $\Lambda_{\rm eff} > 0$.

Now we show that our approximation, $\Delta_p \ll \xi_p + \Phi$, equivalent to $ \delta\alpha_p \ll 1$, is consistent as long as the condensate is dilute. To see that, consider the ansatz~\eqref{eq:the-ansatz} and use the approximation, $\Delta_p \ll \xi_p + \Phi$, to write the density~\eqref{eq:density} as
\begin{equation}
n \approx \frac{p_0}{8\pi}\delta^2\int \alpha_p^2 dp,
\label{eq:density-dilute}
\end{equation}
where  $\delta\ll1$, as appropriate for the onset of superconductivity, and $\alpha_p$ is a normalizable bound state wavefunction of the single pair Schr\"odinger equation \eqref{eq:Sch_pair_Coulomb}. This implies that $\alpha_p$ is peaked around $p_0$, and has a finite width proportional to the inverse of the pair-size, $\zeta$. Hence, one can estimate $\int\alpha_p^2 dp \sim \alpha_{p_0}^2 / \zeta$. Applying this to Eq.~\eqref{eq:density-dilute}, we conclude that in order for the ansatz Eq.~\eqref{eq:the-ansatz} to be valid, we must have
\begin{equation}
\delta^2\alpha_{p_0}^2 \sim \frac{n\zeta}{p_0} \ll 1.
\label{eq:approx-criterion}
\end{equation}

\begin{figure}[h]
\includegraphics[width=\columnwidth]{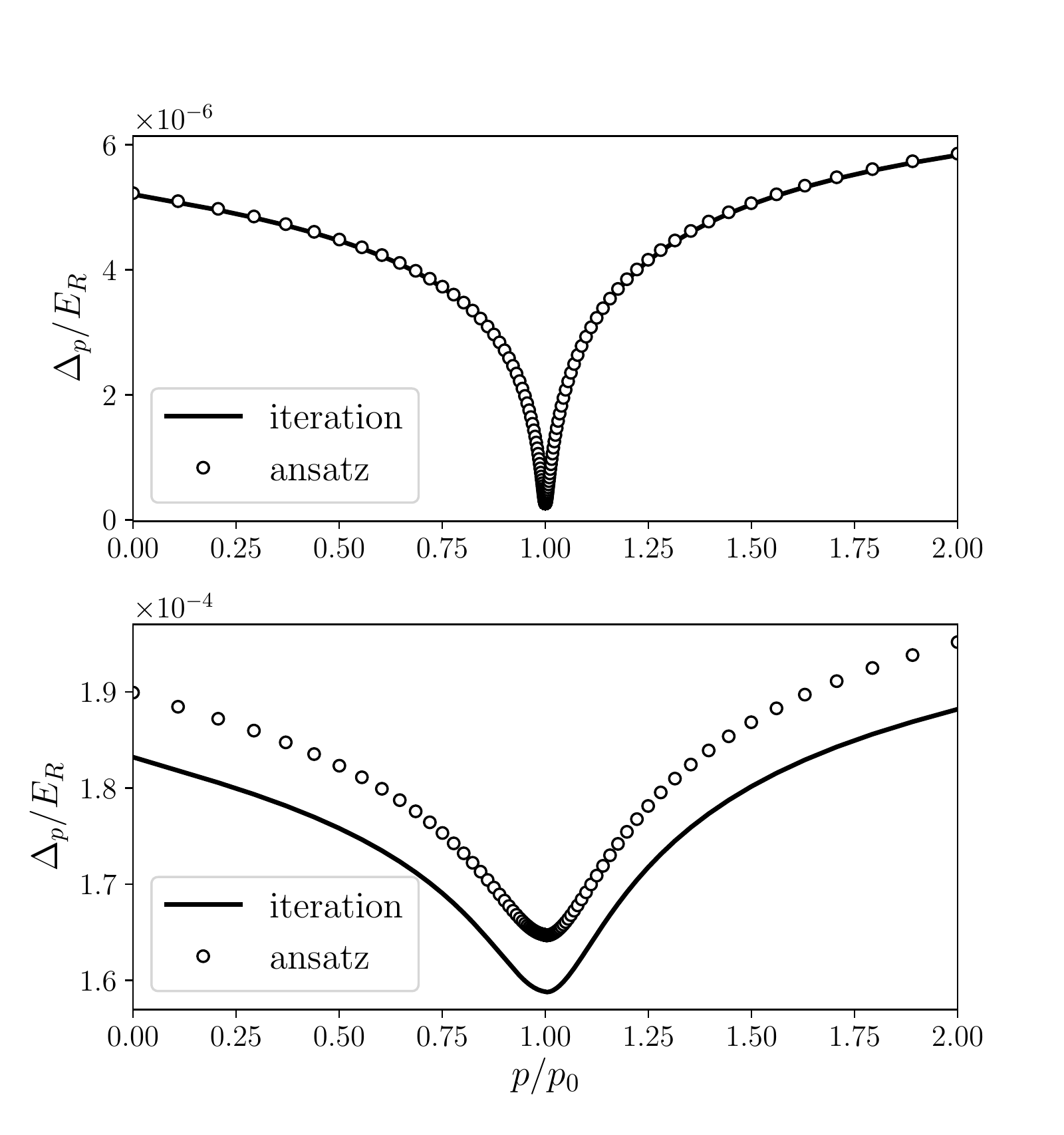}
\caption{Comparing numerical solution of the Eq.~\eqref{eq:self-consis-colmb-1d} (labeled as {\em iteration}) and the ansatz~\eqref{eq:the-ansatz} (labeled as {\em ansatz}) for $\Lambda = 0$ (top panel) and $\Lambda = E_R/100$ (bottom panel). $g$ is chosen to be slightly above $g_c$, such that the diluteness condition~\eqref{eq:approx-criterion} holds. For the top panel, $g/\kappa~=~0.175$, and for the bottom panel, $g/\kappa = 0.52$. In both cases screening is very weak: $dp_0 = 1000$ and $E_c = E_R/10$. In order to make the ansatz solution, the Schr\"odinger equation~\eqref{eq:Sch_pair_Coulomb} has been solved numerically to find $\alpha_p$ and then using Eq.~\eqref{eq:Phi-E} and \eqref{eq:Phi_eq} the normalization constant $\delta$ was found.}
\label{fig:ansatz-iteration}
\end{figure}

At the point of transition, $g = g_c$, the bound state energy of a single pair, as well as density-dependent quantities $\Phi$ and $n$, are zero, while the pair has a finite size for $\Lambda > 0$, which means \eqref{eq:approx-criterion} is valid at the transition point. As $g$ increases above $g_c$ the HF field $\Phi$ and the density $n$ both increase with $g - g_c$, while the pair-size slowly shrinks. Therefore, we conclude that in the case of $\Lambda>0$, one can think of the superconducting state as a dilute condensate of well-defined pairs as long as the system is in the vicinity of the transition point.
Fig.~\ref{fig:ansatz-iteration} shows that the solution constructed by the ansatz~\eqref{eq:the-ansatz} matches the numerical solution of Eq.~\ref{eq:self-consis-colmb-1d} very well.

Some insight into the shape of the $g_c(\Lambda)$ line for the superconductor-insulator transition on the $\Lambda>0$ side of the phase diagram can be deduced from Eqs.~\eqref{eq:Ebquadratic} and~\eqref{eq:Eblinear}. Setting $E_b=0$ in those equations, we see that in the vicinity of $\Lambda=0$ the transition line has a finite slope, $g_c-g_{cb}\propto \Lambda$, while for $g_c-g_{cb}\gtrsim g_{cb}$ we have $g_c-g_{cb}\propto \sqrt{\Lambda}$. Both of these observations are confirmed numerically in Fig.~\ref{fig:PDcoulomb}.

Next we consider the case $\Lambda <0$, when the bare chemical potential is inside the band. This situation is illustrated in the insert of Figure~\ref{fig:bands}. For $g=0$, one has a gas of interacting electrons, with density strongly reduced from the nominal value for the non-interacting case due to the strong electrostatic effects. We can determine the normal-state density by noting that since $\xi_p + \Phi = \kappa (p-p_0)^2 - |\Lambda+ \Phi|$ in Eq.~\eqref{eq:density} changes sign as a function of $p$, there is a finite electron density even in the absence of superconductivity, $\Delta_p = 0$.  For $d\gg \zeta_0$, solving Eqs.~\eqref{eq:Phi_eq} and~\eqref{eq:density} together leads to
\begin{equation}
|\Lambda + \Phi| \approx \frac{\zeta_{0}^2}{d^2}\frac{\Lambda^2}{E_R} \ll |\Lambda|.
\label{eq:L+P}
\end{equation}
This shows that negative bare $\Lambda$ causes positive $\Phi$, the magnitude of which is comparable to $|\Lambda|$, so that the renormalized Fermi energy of the electrons counted from the bottom of the band, $|\Lambda + \Phi|$, is  much smaller than the bare one, $|\Lambda|$. The Coulomb repulsion pulls the effective chemical potential down (decreases $|\Lambda + \Phi|$) so as to reduce the density of electrons. We can define a normal-state Fermi momentum, counted from $p_0$, associated with the normal state Fermi energy, $|\Lambda + \Phi|$:
\begin{align}
    p_F=p_0\sqrt{\frac{|\Lambda + \Phi|}{E_R}}\approx \frac{p_0 \zeta_0}{d}\frac{|\Lambda|}{E_R}.
    \label{eq:pf}
\end{align}

\begin{figure}[h]
\includegraphics[width=\columnwidth]{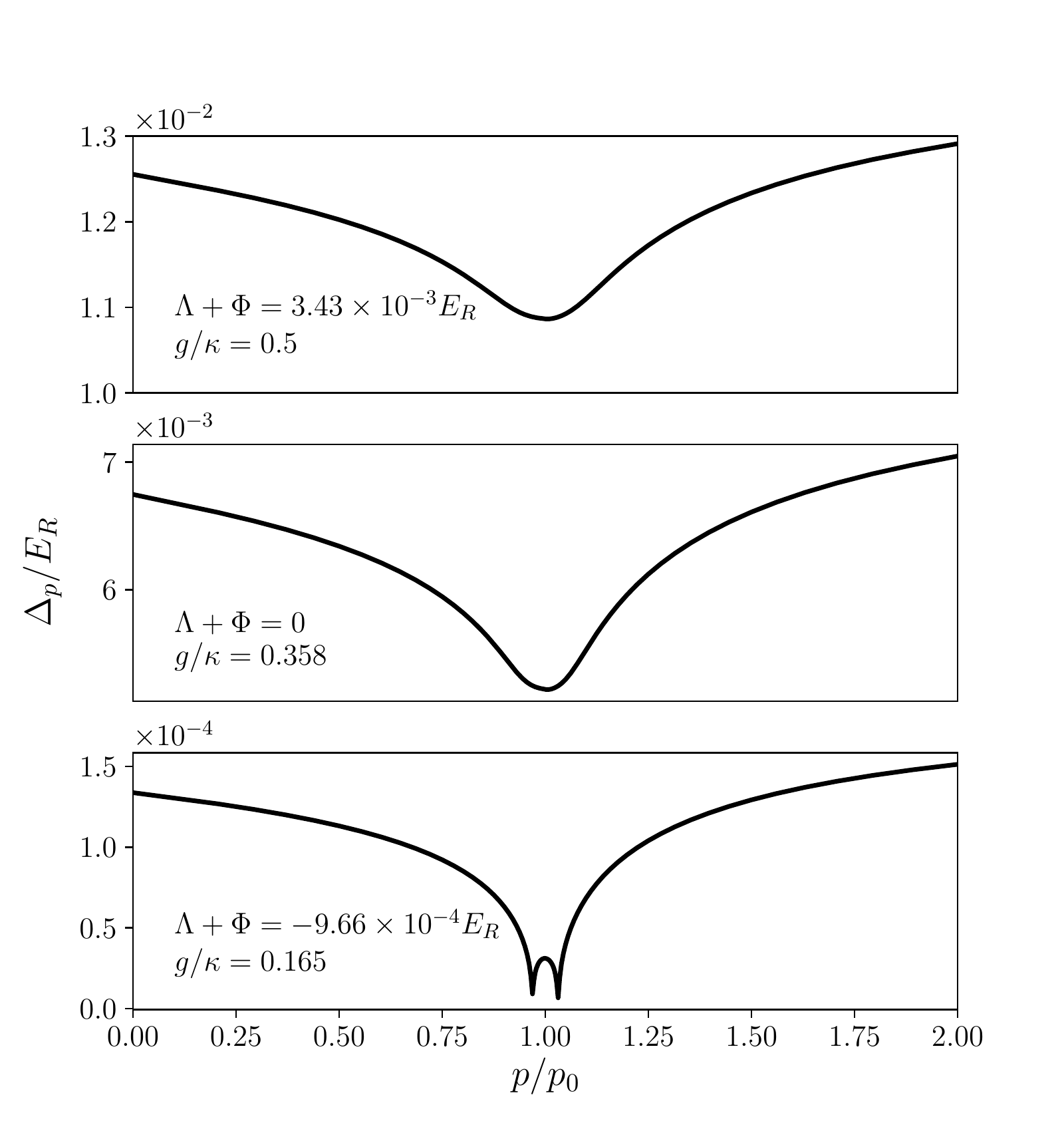}
\caption{Typical $\Delta_p$ numerical solution of Eq.~\ref{eq:self-consis-colmb-1d} for $\Lambda=-E_R$ and very weakly screened Coulomb: $dp_0 = 1000$ and $E_c = E_R /10$. The lowest panel is a typical solution of the BCS kind while the top one was found in the BEC region. The middle panel represents the boundary case where $\Lambda + \Phi = 0$. For each panel numerical values of $g$ and $\Lambda + \Phi$ is written on the plot.}
\label{fig:BCS-BEC}
\end{figure}

Qualitative features of the superconducting transition on the $\Lambda<0$ side depend on the relation between $p_F$ and $1/\zeta_0$, since $1/\zeta_0$ determines the spread of momenta around $p_0$ that participate in the formation of the two-particle bound state. For $p_F\lesssim 1/\zeta_0$, the transition is strongly affected by the formation of the bound state on the BEC side. In that case, one expects $\Delta_p$ with a strong momentum dependence, whose form is largely similar to the one obtained for $\Lambda=0$ at large momenta, $|p-p_0|\gtrsim p_F$ (see Fig.~\ref{fig:BCS-BEC}), with some modifications around the normal-state Fermi surface. In this regime, we expect a value slightly less than $g_{cb}$ for $g_c$, the reduction being due to the increased right hand side of the self-consistency equation -- the first of Eqs.~\eqref{eq:self-consis-colmb-1d} -- in the presence of zeros of $\xi_p$. For $p_0\gg p_F\gtrsim 1/\zeta_0$, the states that participated in the formation of the bound state are completely Pauli-blocked, and BEC physics is irrelevant. In this limit, one expects the pairing physics to be dominated by the Fermi surface, with only the value of $\Delta_p$ at the Fermi momentum being important. Since we are interested in the BEC-BCS crossover physics, we do not pursue the problem of determining the normal-superconducting phase boundary in full rigor, and present only numerical results here, see Fig.~\ref{fig:PDcoulomb}. Of importance to us is the undoubtful fact that the BCS phase does exist on the $\Lambda<0$ side, and the critical $g_c(\Lambda)$ separating the normal and superconducting phases lies below the BEC-BCS crossover line. This latter line will be described below, and is precisely determined numerically.

To see how the $\Delta_p$ solutions of Eq.~\ref{eq:self-consis-colmb-1d} behave on the $\Lambda < 0$ side, we present a typical numerical solution for $\Lambda = - E_R$ in Fig.~\ref{fig:BCS-BEC}. The lowest panel corresponds to the smallest $g$ where $\Lambda + \Phi < 0$ and we are in the BCS region; the middle panel belongs to the boundary case where $\Lambda + \Phi = 0$, which corresponds to the BEC-BCS crossover, see below; and finally for an even larger $g$ in the top panel $\Lambda + \Phi$ becomes positive and we are in the BEC region. Notice how in the BCS region $\Delta_p$ has two very sharp dips on the Fermi momenta and as we go into the BEC region it becomes more like the solutions on the positive side of $\Lambda$ presented in Fig.~\ref{fig:ansatz-iteration}.

It is also noteworthy that we are still within the validity of the moat-band model even for $\Lambda=-E_R$, since the normal-state doping level is much smaller than $|\Lambda|$, see Eq.~\eqref{eq:L+P}. This tendency is seen in Fig.~\ref{fig:BCS-BEC}: while the bare value of $\Lambda = - E_R$ in all three cases shown there, the renormalized values of $|\Lambda + \Phi|$ are indeed much smaller than $|\Lambda|$, and are in agreement with Eq.\eqref{eq:L+P}.

The BEC-BCS crossover line can be deduced from the same considerations as in the case of neutral fermions, except the condition $\Lambda=0$ should be replaced with $\Lambda+\Phi(g,\Lambda)=0$, which implicitly defines a line in the $(g,\Lambda)$ plane. It is clear that this line exists only on the $\Lambda\leq0$ side of the phase diagram, and starts from point $(g_{cb},0)$. The rest of the line can be obtained numerically in the following way: first, one sets $\Lambda+\Phi=0$ in the equation for $\Delta_p$, and thus obtains a solution for every $g>g_{cb}$. Having obtained a solution for $\Delta_p$ for a particular $g$, one then substitutes this solution into the equation for $\Phi$, Eq.~\eqref{eq:Phi_eq}, in which $\Phi$ itself in the left hand side is replaced with $|\Lambda|$. As a result, one obtains $|\Lambda(g)|$ that defines the BEC-BCS crossover line. The result of implementing this program is shown in Fig,~\ref{fig:PDcoulomb} as a dashed line on the $\Lambda_{\rm eff} < 0$ side. This concludes the construction of the phase diagram for the case of charged fermions.

We leave the comprehensive study of the model at finite temperature for future work.
Here we would like to provide a simple estimate of the magnitude of the temperature at which superconductivity develops.
The critical temperature of a two-dimensional superconductor is controlled by the Berezinskii-Kosterlitz-Thouless (BKT) physics which describes binding of vortex-antivortex pairs of the superconducting phase.
Typically, the BKT transition temperature $T_{\rm BKT}$, which describes the onset of the two-dimensional phase coherence, is much lower than the mean-field theory result $T_{\rm mf}$,
roughly set by the magnitude of the superconducting order parameter. In case of neutral fermions with moat dispersion and no Coulomb repulsion Ref.~\onlinecite{bcs-bec-soc} finds that
$k_B T_{\rm BKT} = \hbar^2 \pi n / (4m_{\rm B})$. Here $n$ is the superconducting density, which controls superfluid stiffness, and $m_{\rm B}$ is the effective mass of the pairs, which approaches to $4 m$ in the limit of large SOC, where $m = 1/2\kappa$ in our notation. For an estimate, we assume a favorable set of parameters: material with large SOC, $E_R = 200 {\rm meV}$ \cite{soc_exp_3}, small effective mass
$m = 0.05 m_e$, and moderate attractive interaction $g/\kappa = 0.1$. For $\Lambda = 0$ numerical solution of \eqref{eq:self-consis-colmb-1d} and \eqref{eq:density} gives
$n \approx 0.004 k_0^2$, where the moat wave vector $k_0 = \sqrt{2 m E_R}/\hbar \approx  5.1\times 10^8 {\rm m}^{-1}$. This leads to a large critical temperature
$T_{\rm BKT} = \pi n E_R/(8k_B k_0^2) \approx 3.5$ Kelvin.
As discussed in the bulk of the paper, Coulomb interaction between electrons strongly suppresses superfluid density $n$. Indeed, turning on finite $E_c = e^2 k_0/(2 \epsilon)$, even with a large
$\epsilon = 100\epsilon_0$ \cite{Goldman2014} and intermediate distance to the gate $d k_0 =1$, strongly reduces the density. We find $n \approx 9\times 10^{-5} k_0^2$ and $T_{\rm BKT} \approx 0.08$ Kelvin. Increasing distance to the gate to  $d k_0 =10$ suppresses superfluid density completely, $n=0$, and removes the superconducting state altogether. This simple estimate shows once again the importance of the proper treatment of the weakly screened electrostatic repulsion in the problem of dilute two-dimensional electron gas with moat dispersion.

\section{Conclusion \label{sec:conclusion}}

Two-dimensional electron gas with a moat dispersion represents convenient model system for studying a superconductor - band insulator transition in the limit of low density of charged carriers. Moat dispersion leads to the enhanced density of states which strongly amplifies effects of pairing and repulsive interactions.

BEC-BCS crossover as a function of the strength of attractive interaction $g$
on the negative-$\Lambda$ axis represents one of the key results of our study. Another qualitatively novel feature is the appearance of the bound electron pairs of {\em finite size}, Eq.\eqref{eq:zeta0}, driven by the competition between local attractive $g$ and long-ranged repulsive Coulomb interaction.

While we model the attraction phenomenologically as a contact-like interaction of BCS type, several recent studies have pointed out that in the metals with strong spin-orbit coupling the attraction may be induced by the spin-orbit interaction itself \cite{Mahmoodian2018,Gindikin2018}. We leave investigation of this interesting possibility for the future.

Our study highlights importance of electrostatic contributions to the electron dispersion in the low density limit. Our findings are of relevance to two-dimensional superconducting heterostructures and surface states where spin-orbit interaction is particularly prominent.

\begin{acknowledgments}
We thank Eugene Mishchenko for numerous insightful discussions at the beginning of this investigation.
This work is supported by the NSF grants DMR-1507054 (H.A. and O.A.S.) and DMR-1853048 (DAP).
\end{acknowledgments}

\appendix
\section{Coulomb kernel, and the effective potential in ``real'' space}\label{appendix:potential}

In this Appendix, we derive the approximate expression for the Coulomb kernel, Eq.~\eqref{eq:ReducedCoulombKernel}, and the effective 1D potential energy for the effective 1D motion that describes the radial problem in the SPWF problem, Eq.~\eqref{eq:potential}.

The full form of the Coulomb kernel is given in Eq.~\eqref{eq:K-function}. For convenience, we repeat it here:
\begin{equation}\label{eq:appendixK}
{\cal{K}}(q, k) = \frac{E_c}{2\pi}\int_0^{2\pi}\frac{\cos^2(\theta_{\bm q k}/2)(1-e^{-d|{\bm {q-k}}|})}
{|{\bm {q-k}}|}d\theta.
\end{equation}
Since we are interested in ${\cal K}(p)\equiv {\cal K}(p_0+p/2,p_0-p/2)$ with $p\ll p_0$, we can neglect the $\cos^2(\theta_{\bm qk}/2)$ term in the definition of ${\cal K}(q,k)$.

We now calculate ${\cal K}(p)$ for several important values ot its argument, starting with the value of ${\cal K}(0)$. To this end, we set $q,k\to p_0$ in Eq.~\eqref{eq:appendixK}, introduce a new variable $t=\sin(\theta/2)$, such that $d\theta\approx 2dt$,  to obtain
\begin{align}\label{eq:apppendixK0}
    {\cal K}(0)\approx \frac{E_c}{\pi p_0}\int^{1}_{0}dt \frac{1-e^{-2d p_0 t }}{t}\approx \frac{E_c}{2\pi p_0}\ln (4\beta^2 d^2p_0^2),
\end{align}
where $\beta=e^{\gamma}$ with $\gamma\approx 0.58$ being the Euler-Mascheroni constant, and to write the second approximate equality, we neglected $O(e^{-2dp_0})$ terms, which are negligible in the weak screening limit, $dp_0\gg 1$.

We now turn to the behavior of ${\cal K}(p)$ for $1/d\ll p\ll p_0$. In this limit, the exponential term in the definition of ${\cal K}(q,k)$, Eq.~\eqref{eq:appendixK}, can be neglected, and we arrive at
\begin{align}\label{eq:apppendixKp}
    {\cal K}(p)\approx \frac{2E_c}{\pi}\int^{1}_{0}dt \frac{1}{\sqrt{4p_0^2t^2+p^2}} \approx \frac{E_c}{2\pi p_0} \ln \frac{16p_0^2}{p^2}.
\end{align}
The accuracy of the final expression here is $O(p^2/p_0^2)$.

Given the two limits, Eqs.~\eqref{eq:apppendixK0} and~\eqref{eq:apppendixKp}, we arrive at  Eq.~\eqref{eq:CoulombKsmallq} of the main text as an interpolating function that connects them, and serves as a good approximation to the full Coulomb kernel.

We now discuss the effective 1D potential that corresponds to as explained in Section~\ref{sec:SPWFweakscreening}, the effective potential energy corresponds to the Fourier transform of the Coulomb kernel, ${\cal K}(q)$, see Eq.~\eqref{eq:ReducedCoulombKernel}. It is thus given by
\begin{align}
U(x)=\int^{\infty}_{-\infty}\frac{dq}{2\pi} e^{iqx}{\cal K}(p_0+q/2,p_0-q/2).
\end{align}
By integrating by parts, this integral is reduced to
\begin{align}
    U(x)=-\frac{1}{ix}\int^{\infty}_{-\infty}\frac{dq}{2\pi} e^{iqx}\frac{d}{dq}{\cal K}(p_0+q/2,p_0-q/2).
\end{align}
At this point one can substitute ${\cal K}(p_0+q/2,p_0-q/2)\to {\cal K}(q)$, which is now appropriate, since it yields a convergent result. This way we obtain
\begin{align}\label{eq:potential_appendix}
    U(x)=\frac{E_c}{2\pi p_0 |x|}\exp\left(-\frac{2|x|}{\beta d}\right).
\end{align}

\bibliography{references}

\end{document}